\documentclass{article}

\usepackage{arxiv}
\usepackage[utf8]{inputenc} % allow utf-8 input
\usepackage{twemojis}
\usepackage[T1]{fontenc}    % use 8-bit T1 fonts
\usepackage{hyperref}       % hyperlinks
\usepackage{url}            % simple URL typesetting
\usepackage{booktabs}       % professional-quality tables
\usepackage{amsfonts}       % blackboard math symbols
\usepackage{nicefrac}       % compact symbols for 1/2, etc.
\usepackage{microtype}      % microtypography
\usepackage{graphicx}
\usepackage{doi}
\usepackage{amsmath}
\usepackage{comment}
\usepackage{xcolor}
\usepackage{alltt}
\usepackage{caption} 
\usepackage{accsupp}
\usepackage[firstpage=true]{background}

\usepackage{hyperref}
\usepackage{float}
\usepackage{authblk}
\definecolor{indigo}{HTML}{4B69C6}
\definecolor{maroon}{HTML}{AA3732}
\definecolor{dgreen}{HTML}{319332}
\definecolor{tpurple}{HTML}{7A3E9D}
\definecolor{iibrown}{HTML}{9C5D27}
\definecolor{dorange}{HTML}{E8A20C}
\definecolor{igrey}{HTML}{D3D3D4}
\DeclareUnicodeCharacter{2090}{}
\DeclareUnicodeCharacter{211D}{}
\DeclareUnicodeCharacter{03BB}{}
\DeclareUnicodeCharacter{0327}{}
\DeclareUnicodeCharacter{2200}{}
\DeclareUnicodeCharacter{2115}{}
\DeclareUnicodeCharacter{2217}{}
\title{Formalizing Chemical Physics using \\the Lean Theorem Prover}

\author[1]{\normalsize Maxwell P. Bobbin}
\author[1]{\normalsize Samiha Sharlin}
\author[1]{\normalsize Parivash Feyzishendi}
\author[1]{\normalsize An Hong Dang}
\author[1]{\normalsize \authorcr Catherine M. Wraback}
\author[1,2]{\normalsize  Tyler R. Josephson}

\affil[1]{\normalsize Department of Chemical, Biochemical, and Environmental Engineering, University of Maryland Baltimore County, \authorcr
1000 Hilltop Circle, Baltimore, MD 21250
}
\affil[2]{\normalsize Department of Computer Science and Electrical Engineering, University of Maryland Baltimore County, \authorcr
1000 Hilltop Circle, Baltimore, MD 21250}

\renewcommand{\shorttitle}{Formalizing Chemical Physics}

\hypersetup{
pdftitle={LeanChemicalPhysics},
pdfsubject={cs.lo},
pdfauthor={Tyler R. Josephson},
pdfkeywords={Proof assistants, Formal verification, Logic, Adsorption, Thermodynamics, Kinematics},
hidelinks
}
\date{}

\begin{document}

\maketitle

\begin{abstract}

% Chemical theory can be made more rigorous using the \textit{Lean theorem prover}, an interactive theorem prover for complex mathematics.
% We formalize the Langmuir \cite{langmuir1918adsorption} and BET \cite{brunauer_emmett_teller_1938} theories of adsorption, making each scientific premise clear and every step of the derivations explicit. Lean's math library, \texttt{mathlib}, provides formally-verified theorems for infinite geometries series, which are central to BET theory. While writing these proofs, Lean prompts us to include mathematical constraints that were not originally reported. We also illustrate how Lean flexibly enables the reuse of proofs that build on more complex theories through the use of \textit{functions}, \textit{definitions}, and \textit{structures}. Finally, we construct scientific frameworks for interoperable proofs, by creating structures for classical thermodynamics and kinematics, using them to formalize gas law relationships like Boyle’s Law and equations of motion underlying Newtonian mechanics, respectively. This approach can be extended to other fields, enabling the formalization of rich and complex theories in science and engineering.

\par Interactive theorem provers are computer programs that check whether mathematical statements are correct. We show how the mathematics of theories in chemical physics can be written in the language of the Lean theorem prover, allowing chemical theory to be made even more rigorous and providing insight into the mathematics behind a theory. We use Lean to precisely define the assumptions and derivations of the Langmuir \cite{langmuir1918adsorption} and BET \cite{brunauer_emmett_teller_1938} theories of adsorption. We can also go further and create a network of definitions that build off of each other. This allows us to define a common basis for equations of motion or thermodynamics and derive many statements about them, like the kinematic equations of motion or gas laws such as Boyle’s Law. This approach could be extended beyond chemistry, and we propose the creation of a library of formally-proven theories in all fields of science. Furthermore, the rigorous logic of theorem provers complements the generative capabilities of AI models that generate code; we anticipate their integration to be valuable for automating the discovery of new scientific theories.
\end{abstract}

\newcommand\Text{Accepted for publication in Digital Discovery}

\SetBgColor{textcolor}
\SetBgOpacity{1}
\SetBgAngle{90}
\SetBgVshift{0.45\textwidth}
\SetBgScale{1.2}
\SetBgColor{blue}
\SetBgContents{\Text}

\keywords{Proof assistants
\and Formal verification
\and Proof assistant
\and Theorem provers
\and Logic
\and Adsorption
\and Thermodynamics
\and Kinematics
\and Theory
}

\section{Introduction}
Theoretical derivations in the scientific literature are typically written in a semi-formal fashion, and rely on human peer reviewers to catch mistakes. When these theories are implemented in software, the translation from mathematical model to executable code also requires humans to catch errors. This reflects the gap between mathematical equations describing models in science and the software written to encode these  \cite{hinsenComputationalScienceShifting2014}. 
This occurs because the computer doesn't understand relationships among the scientific concepts and mathematical objects under study, it simply executes the code given it. 
Here, we recommend an alternative: interactive theorem provers that enable the mathematics and programming of science to be expressed in a rigorous way, with the logic checked by the computer. %They are also expressive enough to describe almost any mathematical statement. 

\subsection{Theorem provers for chemical theory}
%  Interactive theorem provers are a type of computer program that aid in the derivation of formal proofs through human-machine interaction \cite{hales2008formal}. They provide a way to interact with the current state of the proof, while the computer verifies each step \cite{rudnicki1992overview, wenzel2002isabelle, barras1997coq, gordon1993introduction, nipkow2002isabelle, owre1992pvs}. Although formal methods of proof are difficult to write, they use strict syntax that not only guarantees correctness, but is also machine-readable and interpretable (see Table \ref{Table 1}). 

\par Interactive theorem provers are a type of computer program used for the creation of formal proofs or derivations, which are a sequence of logical deductions used to prove a theorem\footnote{Mathematical terms that may be unfamiliar to the reader, like \textit{theorem}, are defined in the glossary, section \ref{Glossary}} is correct \cite{hales2008formal}. They provide a way to write a proof step by step, while the computer verifies each step is logically correct \cite{rudnicki1992overview, wenzel2002isabelle, barras1997coq, gordon1993introduction, nipkow2002isabelle, owre1992pvs}. Formal proofs are used extensively in mathematics to prove various theories. On the other hand, scientific theory tends to use informal proofs when deriving its theories, since they are easier to write and understand (see Table \ref{Table 1}).
\begin{table}[ht]

\centering
\BeginAccSupp{ActualText={The table comparing hand-written proofs and formal proofs. The left column pertains to hand-written proofs, while the right column relates to formal proofs. Features of hand-written proofs include informal syntax, intended for human readers, potential to exclude information, possible presence of mistakes, the necessity of human proofreading, and the relative ease of writing. In contrast, formal proofs are characterized by a strict, computer language syntax, machine-readability and executability, inability to miss assumptions or steps, rigorous verification by a computer, automated proof checking, and a level of difficulty in writing. The caption of the table reads "Comparison of hand-written and formalized proofs".}}
\begin{tabular}{@{}ll@{}}
\toprule
\textbf{Hand-written proofs} & \textbf{Formal proofs}           \\ \midrule
Informal syntax              & Strict, computer language syntax \\
Only for human readers       & Machine-readable and executable  \\
Might exclude information    & Cannot miss assumptions or steps \\
Might contain mistakes       & Rigorously verified by computer  \\
Requires human to proofread  & Automated proof checking         \\
Easy to write                & Challenging to write             \\ \bottomrule
\end{tabular}
\EndAccSupp{}
\caption{Comparison of hand-written and formalized proofs.}
\label{Table 1}
\end{table}
 
% An interactive theorem prover may appear to be performing a symbolic manipulation of mathematical expressions, like a Computer Algebra System (CAS) such as SymPy \cite{meurer2017sympy}, but such manipulations are only permitted when supported by a proof with axiomatic foundation. For example, whether multiplication is commutative depends on the types of objects being multiplied: a × b = b × a when a and b are scalars, but A × B \(\ne\) B × A when A and B are matrices.  In CAS systems, matrix multiplication is made non-commutative by imposing special conditions \cite{meurer2017sympy} whereas theorem provers only allow changes that are proven to be valid. So, rearranging  a × b to b × a is only possible in theorem provers once their equivalence has been proven true and changing A × B to B × A is not permitted when it has not been proven true. Consequently, theorem provers provide the utmost degree of confidence by allowing all maths to be constructed from the base axioms of mathematics (see Table \ref{Table 2}). 

Scientists are generally familiar with computer algebra systems (CAS) that can symbolically manipulate mathematical expressions (see Table \ref{Table 2}). These systems include SymPy \cite{meurer2017sympy} and Mathematica. These systems are used frequently for scientific applications but come at the cost of being unsound, meaning they can have false conclusions. 

Theorem provers are more rigorous than computer algebra systems, because they require computer-checked proofs before permitting operations, thereby preventing false statements from being proven. For example, $a \times b = b \times a$ is true when $a$ and $b$ are scalars, but $A \times B \neq B \times A$ when $A$ and $B$ are matrices.  CAS impose special conditions to disallow $A \times B = B \times A$ \cite{meurer2017sympy}, whereas theorem provers only allow changes that are proven to be valid. %So, rearranging  a × b to b × a is only possible in theorem provers once their equivalence has been proven true and changing A × B to B × A is not permitted when it has not been proven true. For example, $a$ and only permit operations that are rigorous proved to be correct.
Theorem provers construct all of their math from a small, base kernel of mathematical axioms, requiring computer-checked proofs for for objects constructed from the axioms. Even the most complicated math can be reduced back to that kernel. Since this kernel is small, verifying it by human experts or with other tools is manageable. Then, all higher-level math built and proved from the kernel is just as trustworthy. This contrasts with how CAS represent and introduce mathematics; because proofs are not required when high-level math is introduced, mistakes could enter at any level, and would require humans to catch and debug them \cite{duranMisfortunesTrioMathematicians2014} (see Table \ref{Table 2}).

\begin{table}[ht]
\centering
\BeginAccSupp{ActualText={The table references two distinct types of computational systems used in mathematical and logical operations. Firstly, Interactive Theorem provers such as the ones mentioned in the citations (including Lean and Isabelle), are tools that assist in the creation and checking of formal logic proofs. Secondly, Computer Algebra Systems including Mathematica, Sympy, and MATLAB, as referenced, are software packages that facilitate symbolic mathematics.}}
\begin{tabular}{@{}ll@{}}
\toprule
\textbf{Interactive theorem provers}  & \textbf{Computer algebra systems}  \\ \midrule
Symbolically transform formulae       & Symbolically transform formulae    \\
Only permit correct transformations   & Human-checked correctness          \\
Verification tool                     & Computational tool            \\
Explicit assumptions                  & Hidden assumptions                 \\
Built off a small, trusted, kernel    & Large program with many algorithms \\
\includegraphics[width=5cm]{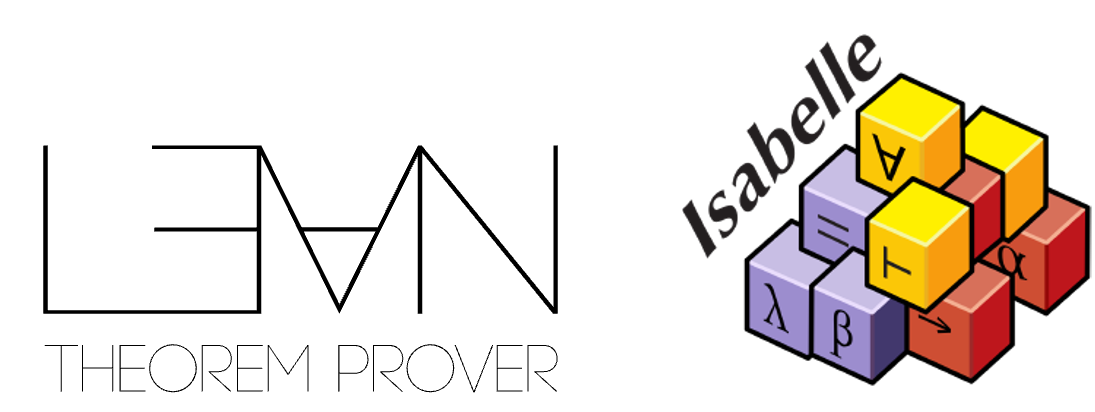} & \includegraphics[width=5cm]{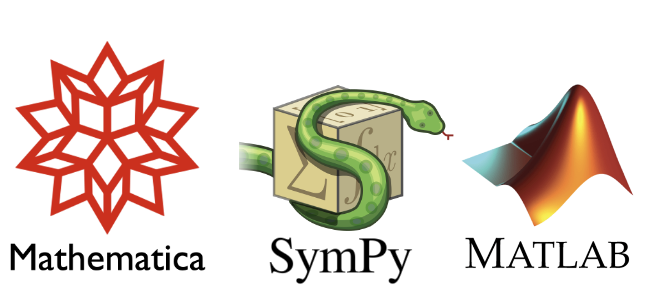}
\\

\bottomrule
\end{tabular}
\EndAccSupp{}
\captionsetup[table]{skip=10pt}
\caption{Interactive Theorem Provers \cite{de_moura_kong_avigad_van_doorn_von_raumer_2015, wenzel2002isabelle} vs. Computer Algebra Systems \cite{wolfram1991mathematica, meurer2017sympy, toolbox1993matlab}.}
\label{Table 2}
\end{table}

Historically, interactive theorem provers have been used to logically connect advanced math theorems to the foundational axioms of mathematics \cite{appel1977,gonthier2008, boldo2015coquelicot, hales2017, buzzard2020, scholzeproof2021}. Before computers, this ``axiomatization'' of mathematics was developed by hand, in works like \emph{Principia Mathematica} by Alfred North Whitehead and Bertrand Russell \cite{1963principia} -- the aim is to write down a minimal list of fundamental assumptions (axioms), and then systematically derive \emph{all} of mathematics from those axioms. Computers play a key role in modern formalization efforts because they can store and verify massive libraries of interconnected theorems collaboratively written by hundreds of mathematicians \cite{hartnettBuildingMathematicalLibrary2020}.

An analogous program to ``axiomatize'' physics was famously articulated as Hilbert's sixth problem \cite{hilbert1902mathematical}. Recent reviews have discussed progress and unsolved questions on this ``endless road'' to describe how all of physics can be derived from a minimal set of axioms \cite{corry1997david,gorban2018hilbert}. Our vision is somewhat distinct from this -- we are inspired by Paleo’s ideas for formalizing physics theories \cite{paleo2012physics} \emph{as a collection of proofs}, instead of aiming to represent science as a single edifice emerging from one set of axioms (though this structure may emerge in the future). In particular, we ask ``How can we formally represent a collection of proofs/derivations using an interactive theorem prover?'' 

Theorem provers have previously been used to formalize derivations in physics: theorems from Newton's \emph{Principia} \cite{fleuriotCombinationNonstandardAnalysis1998, fleuriotProvingNewtonPropositio1999}, versions of relativity theory \cite{stannett2014, lu2017formalization}, electromagnetic optics \cite{khan2014formal}, and geometrical optics \cite{siddique2015formal} have been described and proved using proof assistants. Artificial intelligence tools for scientific discovery have also used theorem provers in designing optical quantum experiments \cite{cervera2021design}, as well as for rediscovering and deriving scientific equations from data and background theory \cite{cornelioCombiningDataTheory2023}.

Here we focus on formalizing fundamental theories in the chemical sciences. Progress toward axiomatizing thermodynamics began with Carath\'eodory in 1909 \cite{Caratheodory1909UntersuchungenD}, with recent developments by Lieb and Yngvason \cite{lieb1999physics}. But broadly, these questions have not been addressed using theorem provers to check the mathematics, which have seen limited use in the chemical sciences. One notable application by Bohrer \cite{Bohrer22} uses a proof assistant that reasons about differential equations and control algorithms \cite{alur2011formal} to describe and prove properties of chemical reactors.
%For instance, the laws of thermodynamics and even more fundamentally proving that macroscopic observables, such as pressure and viscosity, can be derived from ensemble averages of microscopic measurements at equilibrium can help to prove that end-to-end implementation of a molecular dynamics simulation conserves energy in the limit of $\Delta$t → 0 and satisfies ergodicity in the limit of t → $\infty$ \cite{frenkel2001}.  Such an approach provides a strong assurance that a code is correct than relying on diagnostics from simulations \cite{merz2018testing}.

\subsection{The Lean theorem prover}

We have selected the Lean theorem prover \cite{avigad2015theorem} for its power as an interactive theorem prover, the coverage of its mathematics library, \texttt{mathlib} \cite{mathlib2020}, and the supportive online community of Lean enthusiasts \cite{zulipLean} with an aim to formalize the entire undergraduate math curriculum \cite{hartnettBuildingMathematicalLibrary2020, UGLean}. Interesting projects in modern mathematics have emerged from its foundations, including Perfectoid Spaces \cite{10.1145/3372885.3373830}, Cap Set Problem \cite{dahmen_et_al2019} and Liquid Tensor \cite{leanliquidrepo} have garnered attention in the media \cite{hartnettProofAssistantMakes2021}. A web-based game, the Natural Number Game \cite{NNG_github}, has been widely successful in introducing newcomers to Lean. As executable code, Lean proofs can be read by language modeling algorithms that find patterns in math proof databases, enabling automated proofs of formal proof statements, including International Math Olympiad problems \cite{proofartifact2021,FormalMath2022}. 

\par We anticipate that Lean is expressive enough to formalize diverse and complex theories across quantum mechanics, fluid mechanics, reaction rate theory,  statistical thermodynamics, and more. Lean gets its power from its ability to define mathematical objects and prove their properties, rather than just assuming premises for the sake of individual proofs. Lean is based on Type Theory \cite{whitehead2012elements, goerss2009simplicial} where both mathematical objects and the relation between them are modeled with types (see Fig.~\ref{leanoverview} in the Supporting Information). Everything in Lean is a term of a \textit{Type}, and Lean checks to make sure that the \textit{Types} match. Natural numbers, real numbers, functions, Booleans, and even proofs are types; examples of terms with these types include the number 1, Euler's number, \(f(x) = x^2\), \texttt{TRUE}, and the proof of BET theory, respectively. Lean is also expressive enough to allow us to define new types, just like mathematicians do  \cite{avigad2015theorem}, which allows us to define specific scientific theories and prove statements about them.

\par In this paper, we show how formalizing chemical theories may look, by demonstrating the tools of Lean through illustrative proofs in the chemical sciences. First, we introduce variables, types, premises, conjectures, and proof steps through a simple derivation of the Langmuir adsorption model. Next, we show how functions and definitions can be used to prove properties of mathematical objects by revising the Langmuir adsorption model through definitions and showing it has zero loading at zero pressure. Finally, we turn to more advanced topics, such as using geometric series to formalize the derivation of the BET equation and using structures to define and prove relationships in thermodynamics and motion.

\section{Methods}

\par Lean has a small kernel, based on dependent type theory \cite{whitehead2012elements, goerss2009simplicial}, with just over 6000 lines of code that allows it to instantialize a version of the Calculus of Inductive Constructions (CoIC) \cite{coquand1986calculus, coquand1988inductively}. The strong normalizing characteristic of the CoIC \cite{coquand1990proof} creates a robust programming language that is consistent. The CoIC creates a constructive foundation for mathematics allowing the entire field of mathematics to be built off of just 6000 lines of code. 

\par In Section \ref{proofs} we outline the proofs formalized using Lean version 3.51.1. We host proofs on a website that provides a semi-interactive platform connecting to the Lean codes in our GitHub repository \href{https://atomslab.github.io/LeanChemicalTheories/} {{\Large\texttwemoji{atom symbol}}}. An extended methods section introducing Lean is in the Supporting Information Section \ref{LeanAdditionalInfo}.

\section{Formalized Proofs}
\label{proofs}

\subsection{Langmuir Adsorption: Introducing Lean Syntax and Proofs} \label{LangmuirProofSection}
We begin with an easy proof to introduce Lean and the concept of formalization. 
The Langmuir adsorption model describes the loading of adsorbates onto a surface under isothermal conditions \cite{langmuir1918adsorption}. Several derivations have been developed \cite{langmuir1918adsorption,volmer1925thermodynamische,masel1996principles,kleman2003soft}; here we consider the original kinetic derivation \cite{langmuir1918adsorption}. First, we present a derivation of the Langmuir model given by the Eq.~\ref{Langmuir Model}, in \LaTeX, then transfer this into Lean and rigorously prove it. We also discuss how these proofs can be improved to be more robust. 

The Langmuir model assumes that all sites are thermodynamically equivalent, the system is at equilibrium, and that adsorption and desorption rates are first order. The adsorption and desorption rates are given by Eq. \ref{Langmuir Adsorption} and Eq. \ref{Langmuir Desorption}, respectively. 
\begin{equation}
    r_{ad} = k_{ad}p_A[S]
    \label{Langmuir Adsorption}
\end{equation}
The symbols \(r_{ad}\), \(k_{ad}\), \(p_A\), and \([S]\) represent the rate of adsorption, the adsorption rate constant, the pressure of the adsorbate gas, and the concentration of available sites on the surface, respectively.
\begin{equation}
    r_d = k_d[A_{ad}]
    \label{Langmuir Desorption}
\end{equation}
Where \(r_{d}\) stands for the rate of desorption, \(k_{d}\) signifies the desorption rate constant, and \([A_{ad}]\) represents the concentration of adsorbed molecules. From assumption (2), \(r_{ad} = r_d\), and with some rearrangement, we get Eq. \ref{Langmuir Equilibrium}.
\begin{equation}
    [S] = \frac {k_d[A_{ad}]}{k_{ad}p_A}
    \label{Langmuir Equilibrium}
\end{equation}
Using the site balance \([S_0] = [S] + [A_{ad}]\), where  \([S_0]\) represents the total concentration of available sites, we arrive at Eq. \ref{Langmuir Intermediate}.
\begin{equation}
    [S_0] = \frac {[A_{ad}]} {\frac{k_{ad}}{k_d}p_A} + [A_{ad}]
    \label{Langmuir Intermediate}
\end{equation}
We can rearrange Eq. \ref{Langmuir Intermediate} into a familiar form, Eq. \ref{Langmuir Intermediate 2}.\footnote{The manuscript we first submitted for peer review included a typo in Eq.~\ref{Langmuir Intermediate 2}, with $[S_0]$ appearing as $[S]$. Neither the authors nor the peer reviewers detected this; it was identified by a community member who accessed the paper on arXiv. Of course, Lean catches such typos immediately.}
\begin{equation}
    \frac{[A_{ad}]}{[S_0]}= \frac{\frac{k_{ad}}{k_d}p_A}{1 + \frac{k_{ad}}{k_d}p_A}
    \label{Langmuir Intermediate 2}
\end{equation}
Using the definition of the fraction of adsorption, \(\theta = \frac {[A_{ad}]}{[S_0]}\), and the definition of the equilibrium constant, \(K_{eq}^A = \frac{k_{ad}}{k_d}\), we arrive at the familiar Langmuir Model, 
Eq.~\ref{Langmuir Model}. 
\begin{equation}
    \label{Langmuir Model}
    \theta _{A}=\frac{K_{eq} p_{A}}{1+K_{eq} p_{A}}
\end{equation}

\par This informal proof is done in natural language, and it doesn't explicitly make clear which equations are premises to the proof and which are intermediate steps. While the key steps from the premises to the conclusion are shown, the fine details of the algebra are excluded. In contrast, Lean requires premises and conjecture to be precisely defined and requires that each rearrangement and cancellation is shown or performed computationally using a tactic. The next part shows how this proof is translated into Lean.
\begin{figure}[H]
\centering
\BeginAccSupp{method=plain,Alt={Screenshot of Langmuir's adsorption model formalization in Lean using VSCode. The left panel displays the ``Code Window'' and the right panel shows the ``Tactic State'' displaying variables and goals. The proof progresses by applying tactics, updating hypotheses and goals at each numbered location in the ``Code Window'', highlighted by the cursor. The turnstile symbol denotes the goal state. Steps 1 and 2 rewrite the adsorption and desorption equations, affirming equilibrium. Steps 3-5 rearrange the variables to match the goal state. The completion of the proof is indicated by a party emoji.}}%
\includegraphics[width=1\textwidth]{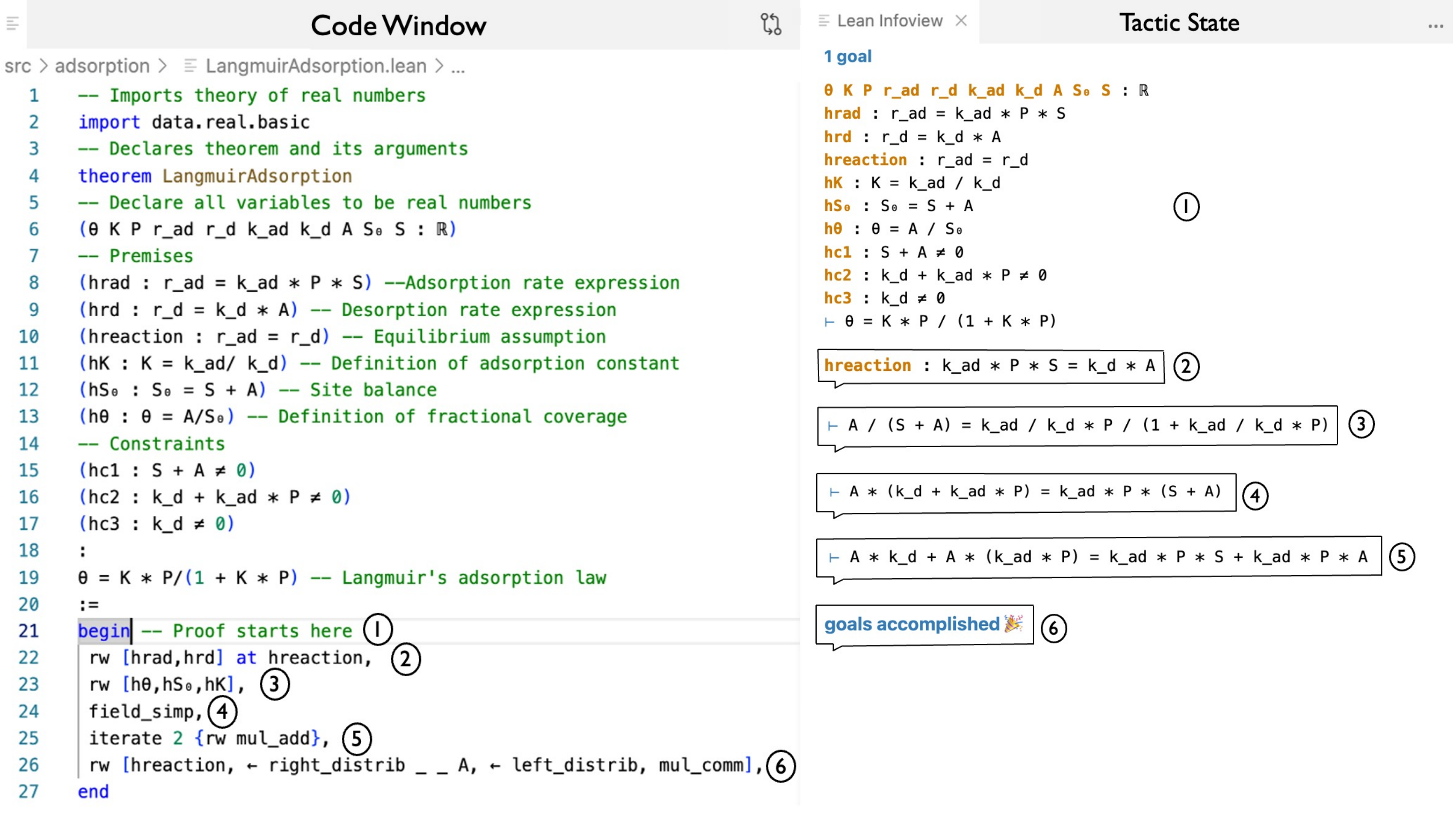}%
\EndAccSupp{} 
\caption{A formalization of Langmuir’s adsorption model, shown as screenshots from Lean operating in VSCode. The left side of the figure shows the “Code Window,” while the right side shows variables and goals at each step in the “Tactic State”. When the user places the cursor at one of the numbered locations in the “Code Window,” VSCode displays the “Tactic State” of the proof. Lean allows the use of Unicode symbols, so we use “$S_{0}$” to represent the total concentration of adsorption sites without needing underscores. The turnstile symbol represents the state of the goal after each step.
As each tactic is applied, hypotheses and/or goals are updated in the tactic state as the proof proceeds. For clarity, we only show the hypothesis that changes after a tactic is applied and how that changes the goal. As an example, the goal state is
the same in steps 1 and 2 since the first tactic rewrites (rw) the equation of adsorption (hrad) and desorption (hrd) into the premise that equilibrium (hreaction) exists. Next, we rewrite (rw), simplify (field\_simp), and otherwise rearrange the variables to exactly equal the goal state (steps 3-5). When the proof is finished, a celebratory message and party emoji appears (6).}
\label{LangmuirProof}
\end{figure}

As shown in Figure \ref{LangmuirProof}, every premise must be explicitly stated in Lean, along with the final conjecture and proof tactics used to show that the conjecture follows from the premises. 
Lean is an \emph{interactive} theorem prover, meaning that the user is primarily responsible for setting up the theorem and writing the proof steps, while Lean continuously checks the work and provides feedback to the user.
The central premises of the proof are expressions of adsorption rate (\textit{hrad}), desorption rate (\textit{hrd}), the equilibrium relation (\textit{hreaction}), and the adsorption site balance (\textit{h$S_{0}$}).
Additional premises include the definition of adsorption constant (\textit{hK}) and surface coverage (\textit{h$\theta$}) from the first four premises, as well as mathematical constraints (\textit{hc1}, \textit{hc2}, and \textit{hc3}) that appear during the formalization. The model assumes the system is in equilibrium, so the adsorption rate, \(r_{ad} = k\_ad*Pₐ*S\) and desorption rate,  \(r_d = k\_d*A\) are equal to each other, where \(k_{ad}\) and \(k_d\) are the adsorption and desorption rate constant respectively, $S$ is concentration of empty sites, and $A$ is the concentration of sites occupied by $A$. After \texttt{begin}, a sequence of tactics rearranges the goal state until the conjecture is proved. Note when performing division, Lean is particular to require that the denominator terms are nonzero. 

\par An interesting part of the proof is that only certain variables or their combinations are required to be not zero. When building this proof, Lean imports the real numbers and the formalized theorems and tactics for them in \texttt{mathlib}. Lean does not permit division by zero, and it will flag issues when a number is divided by another number that could be zero. Consequently, we must include additional hypotheses \textit{hc1}-\textit{hc3} in order to complete the proof. These provide the minimum mathematical requirements for the proof; more strict constraints requiring rate constants and concentrations to be positive would also suffice. These ambiguities are better addressed by using \textit{definitions} and \textit{structures}, which enable us to prove properties about the object. Nonetheless, this version of the Langmuir proof is still a machine-readable, executable, formalized proof.

\par Though this is a natural way to write the proof, we can condense the premises by using local definitions. For instance, the first two premises \textit{hrad} and \textit{hrd} can be written into \textit{hreaction} to yield \textit{k\_ad*Pₐ*S = k\_d*A} and we can also write expressions of \textit{h\(\theta\)} and \textit{hK} in the goal statement. While \textit{hrad}, \textit{hrd}, \textit{h\(\theta\)}, and \textit{hK} each have scientific significance, in this proof, they are just combinations of real numbers. Alternative versions of this proof are described in SI Section \ref{LangmuirSI}.

\subsection{Langmuir Revisited: Introducing Functions and Definitions in Lean}
Functions in Lean are similar to functions in imperative programming languages like Python and C, in that they take in arguments and map them to outputs. However, functions in Lean (like everything in Lean) are also objects with properties that can be formally proved. 
\par Formally, a function is defined as a mapping of one set (the domain) to another set (the co-domain). The notation for a function is given by the arrow "$\rightarrow$". For instance, the function, conventionally written as $Y = f(X)$ or $Y(X)$, maps from set \textit{X} to set \textit{Y} is written as $X \rightarrow Y$ in arrow notion.\footnote{These types are easily extended to functionals, which are central to density functional theory. A function that takes a function as an input can be defined by \((\mathbb{R} \rightarrow \mathbb{R}) \rightarrow \mathbb{R}\)}.

Importantly, the arrow "\(\rightarrow\)" is also used to represent the conditional statement (if-then) in logic, but this is not a duplication of syntax. Because everything is a term of \textit{Type} in Lean, functions map type \textit{X} to type \textit{Y}; when each type is a proposition, the resulting function is an if-then statement.

\par As stated in the introduction, Lean's power comes from the ability to define objects globally, not just postulate them for the purpose of local proof. When a mathematical object is formally defined in Lean, multiple theorems can be written about it with certainty that all proofs pertain to the same object. In Lean, we use \textit{def} to define new objects and then prove statements about these objects. The \textit{def} command has three parts:  the arguments it takes in (the properties of the object),  the type of the output, and the proof that the object has such a type. In Lean:
\begin{alltt}
    \textcolor{indigo}{def} \textcolor{maroon}{name} properties : \textcolor{tpurple}{type} := proof of that type
\end{alltt}
For instance, we can define a function that doubles a natural number:
\begin{alltt}
    \textcolor{indigo}{def} \textcolor{maroon}{double} : \textcolor{tpurple}{\(\mathbb{N}\)} \(\rightarrow\) \textcolor{tpurple}{\(\mathbb{N}\)} := \(\lambda\) n : \textcolor{tpurple}{\(\mathbb{N}\)}, n + n
\end{alltt}
The \(\lambda\) symbol comes from lambda calculus and is how an explicit function is defined. After the lambda symbol is the variable of the function, \textit{n} with type \(\mathbb{N}\). After the comma is the actual function. By hand, we would write this as \(f(n) = n + n\). This function doubles any natural number, as the name suggests. We could use it, for example, to show:
\begin{alltt}
    \textcolor{maroon}{double} \textcolor{indigo}{(}3 : \textcolor{tpurple}{\(\mathbb{N}\)}\textcolor{indigo}{)} = \textcolor{indigo}{(}6 : \textcolor{tpurple}{\(\mathbb{N}\)}\textcolor{indigo}{)}
\end{alltt}
\par In the previous section, we showed an easy-to-read derivation of Langmuir adsorption, and in SI Section \ref{LangmuirSI}, we improved the proof using local definitions. Here, we improve it further by defining the Langmuir model as an object in Lean and then showing the kinetic derivation of that object. This way, the object defining the single-site Langmuir model can be reused in subsequent proofs, and all are certain to refer to the same object. 

\par We define the model as a function that takes in pressure as a variable. Given a pressure value, the function will compute the fractional occupancy of the adsorption sites. In Lean, this looks like \href{https://atomslab.github.io/LeanChemicalTheories/adsorption/langmuir_kinetics.html#langmuir_single_site_model}{{\Large\texttwemoji{atom symbol}}}:
\begin{alltt}
    \textcolor{indigo}{def} \textcolor{maroon}{langmuir_single_site_model} \textcolor{indigo}{(}equilibrium_constant : \textcolor{tpurple}{\(\mathbb{R}\)}\textcolor{indigo}{)} : \textcolor{tpurple}{\(\mathbb{R}\)} → \textcolor{tpurple}{\(\mathbb{R}\)} := 
    \(\lambda\) P : \textcolor{tpurple}{\(\mathbb{R}\)}, equilibrium_constant*P/\textcolor{indigo}{(}1+equilibrium_constant*P\textcolor{indigo}{)}
\end{alltt}
The \(\lambda\) symbol comes from \(\lambda\)-calculus \cite{barendregt2013lambda} and is one way to construct functions. It declares that P is a real number that can be specified. When the real number is specified, it will take the place of P in the equation. The definition also requires the equilibrium constant to be specified\footnote{This definition can be specified in multiple ways. The pressure could be required as an input like the equilibrium constant, or the equilibrium constant can be specified as a variable in the function like pressure. Any of these definitions work, and it is possible to prove congruence between them. We chose this way to purposefully show both definitions and functions in Lean.}. 
\par With this, the kinetic derivation of Langmuir can be set up in Lean like this \href{https://atomslab.github.io/LeanChemicalTheories/adsorption/langmuir_kinetics.html#langmuir_single_site_kinetic_derivation}{{\Large\texttwemoji{atom symbol}}}:
\begin{alltt}
    \textcolor{indigo}{theorem} \textcolor{maroon}{langmuir_single_site_kinetic_derivation}
    \textcolor{indigo}{\{}Pₐ k_ad k_d A S : \textcolor{tpurple}{\(\mathbb{R}\)}\textcolor{indigo}{\}}
    \textcolor{indigo}{(}hreaction : let r_ad := k_ad*Pₐ*S, r_d := k_d*A in r_ad = r_d\textcolor{indigo}{)}
    \textcolor{indigo}{(}hS : S \(\ne\) 0\textcolor{indigo}{)}
    \textcolor{indigo}{(}hk_d : k_d \(\ne\) 0\textcolor{indigo}{)}
    :
    \textcolor{indigo}{let} \(\theta\) := A/\textcolor{indigo}{(}S+A\textcolor{indigo}{)},
        K := k_ad/k_d \textcolor{indigo}{in}
        \(\theta\) = \textcolor{maroon}{langmuir_single_site_model} K Pₐ :=
\end{alltt}
This derivation is almost exactly like the proof in SI Section \ref{LangmuirSI}; the only difference is the use of the Langmuir model as an object. After the \texttt{langmuir\_single\_site\_model} simplifies to the Langmuir equation, the proof steps are the same. 

Using the definition makes it possible to write multiple theorems about the same Langmuir object. We can also prove that the Langmuir expression has zero loading at zero pressure, and in the future we can show that it has a finite loading in the limit of infinite pressure, and converges to Henry's Law in the limit of zero pressure \href{https://atomslab.github.io/LeanChemicalTheories/adsorption/langmuir_kinetics.html#langmuir_zero_loading_at_zero_pressure}{{\Large\texttwemoji{atom symbol}}}. Definitions and structures, as we will see in later sections, are crucial to building a web of interconnected scientific objects and theorems.

\subsection{BET Adsorption: Formalizing a complex proof}
\begin{figure}[H]
\centering
\BeginAccSupp{ActualText={The figure provides a comparison between the Langmuir and BET adsorption models. Unlike Langmuir, the BET model accommodates the formation of infinite layers over previously adsorbed particles. In this context, 'θ' signifies the proportion of the surface that has been adsorbed, 'V' refers to the overall volume adsorbed, 'V0' is the volume of a complete monolayer adsorbed on a unit area, 'si' designates the surface area of each sequential layer, 's0' is the surface area of the initial layer, and 'x' and 'C' are constants associated with the heat of adsorption for the respective molecule layers.}}
\includegraphics[width=1\textwidth]{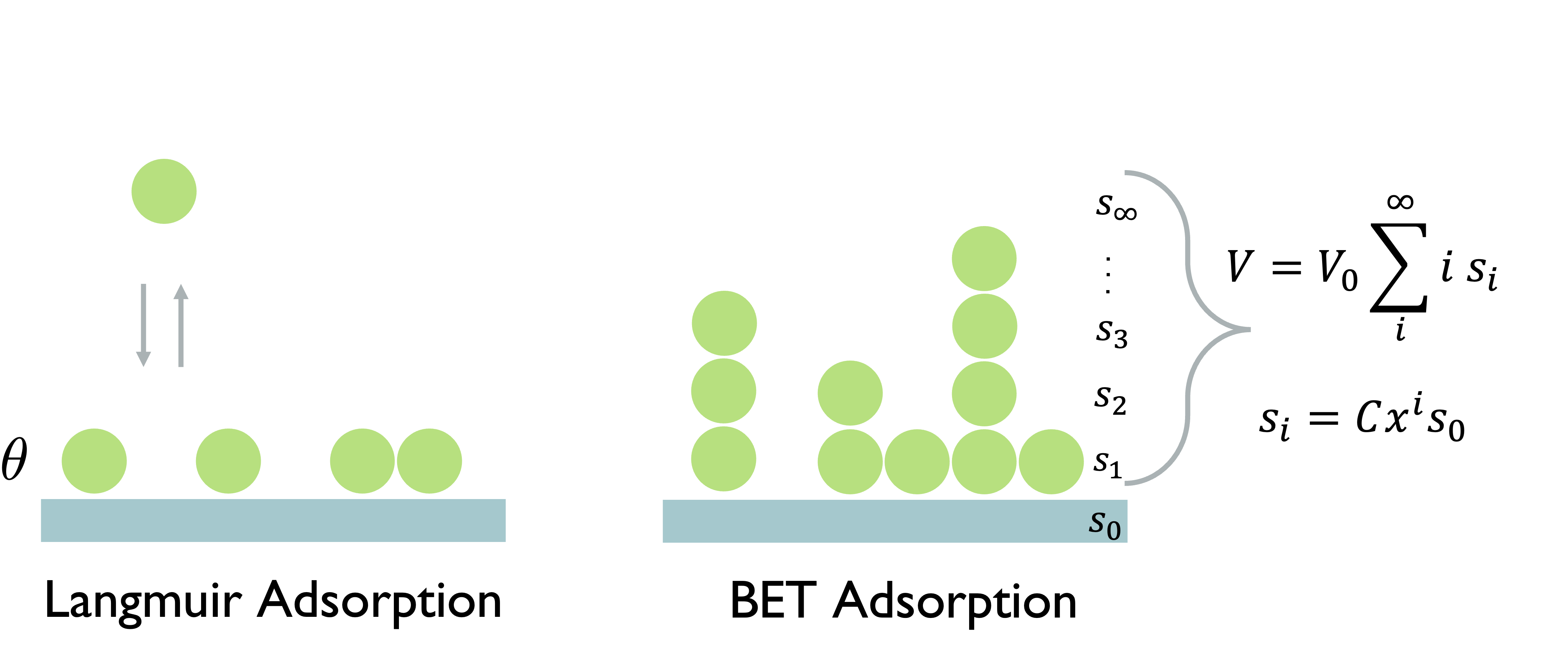}
\EndAccSupp{}
\caption{Langmuir model vs BET model. The BET model, unlike Langmuir, allows particles to create infinite layers on top of previously adsorbed particles. Here $\theta$ is fraction of the surface adsorbed, V is total volume adsorbed, \(V_0\) is the volume of a complete unimolecular layer adsorbed in unit area, \(s_i\) is the surface area of the i\textsuperscript{th} layer, \(s_0\) is the surface area of the zeroth layer, and \textit{x} and \textit{C} are constants that relates heats of adsorption of the molecule in layers.}
\label{LangmuirBET}%
\end{figure}

Brunauer, Emmett, and Teller introduced the BET theory of multilayer adsorption (see Fig.~\ref{LangmuirBET}) in 1938 \cite{brunauer_emmett_teller_1938}. We formalize this derivation, beginning with Equation 26 from the paper, which is shown here in Eq.~\ref{BET 26}:

\begin{equation}
    \label{BET 26}
    \frac{V}{A*V_0} = \frac{Cx}{(1-x)(1-x+Cx)}
\end{equation}

Here A is the total area adsorbed by all (infinite) layers expressed as sum of infinite series:

\begin{equation}
    \label{Catalyst Surface Area}
    A = \sum_{i=0} ^{\infty} s_i =  s_0(1+C\sum_{i=1} ^{\infty} x^i)
\end{equation}

and V is the total volume adsorbed is given by:
\begin{equation}
    \label{Volume Adsorbed}
    V = V_0 \sum_{i=0} ^{\infty} is_i = Cs_0V_0 \sum_{i=1} ^{\infty} ix^i
\end{equation}

The variables \textit{y}, \textit{x} and \textit{C} are expressed in the original paper as shown through Eq.~\ref{BET y} to \ref{BET C}: 
\begin{equation}
    \label{BET y}
    y = PC_1,\; where \;C_1 = (a_1/b_1)e^{E_1/RT}
\end{equation}

\begin{equation}
    \label{BET x}
    x = PC_L,\; where \;C_L = e^{E_L/RT}/g
\end{equation}

\begin{equation}
    \label{BET C}
    C = y/x = C_1/C_L
\end{equation}
where \(a_1\), \(b_1\), and \(g\) are fitted constants, \(E_1\) is the heat of adsorption of the first layer, \(E_L\) is for the second (and higher) layers (also the same as heat of liquefaction of the adsorbate at constant temperature), \(R\) is the universal gas constant, and \(T\) is temperature. In Eq.~\ref{BET y} and \ref{BET x}, everything besides the pressure term is constant, since we are dealing with an isotherm, so we group the constants together into one term.

\par These constants, along with the surface area of the zeroth layer, given by \(s_0\), saturation pressure, and the three constraints are defined using the \textit{constant} declaration in Lean. Mathematical objects can also be defined in other ways such as \textit{def}, \textit{class} or \textit{structure} \cite{avigad2015theorem} but for this proof we will use \textit{constant} which is convenient for such simple objects. We will illustrate later in our thermodynamics proof how constants can be merged into a Lean \textit{structure} for reusability.

In Lean, this is \href{https://atomslab.github.io/LeanChemicalTheories/adsorption/BETInfinite.html#C_L}{{\Large\texttwemoji{atom symbol}}}:

\begin{alltt}
    \textcolor{indigo}{constants} \textcolor{indigo}{(}C_1 C_L s_0 P_0: \textcolor{tpurple}{\(\mathbb{R}\)}\textcolor{indigo}{)}
    \textcolor{indigo}{(}hCL : 0 < C_L\textcolor{indigo}{)} \textcolor{indigo}{(}hC1 : 0 < C_1\textcolor{indigo}{)} \textcolor{indigo}{(}hs_0 : 0 < s_0\textcolor{indigo}{)}
\end{alltt}
\par With these \textit{constant} declarations, we can now define \textit{y, x,} and \textit{C} in Lean  as \href{https://atomslab.github.io/LeanChemicalTheories/adsorption/BETInfinite.html#BET_first_layer_adsoprtion_rate}{{\Large\texttwemoji{atom symbol}}}:

\begin{alltt}
    \textcolor{indigo}{def} \textcolor{maroon}{BET_first_layer_adsorption_rate} \textcolor{indigo}{(}P : \textcolor{tpurple}{\(\mathbb{R}\)}\textcolor{indigo}{)} := \textcolor{indigo}{(}C_1\textcolor{indigo}{)}*P
    \textcolor{indigo}{local notation} `y' := BET_first_layer_adsorption_rate
    
    \textcolor{indigo}{def} \textcolor{maroon}{BET_n_layer_adsorption_rate} \textcolor{indigo}{(}P : \textcolor{tpurple}{\(\mathbb{R}\)}\textcolor{indigo}{)}:= \textcolor{indigo}{(}C_L\textcolor{indigo}{)}*P
    \textcolor{indigo}{local notation} `x' := BET_n_layer_adsorption_rate
    
    \textcolor{indigo}{def} \textcolor{maroon}{BET_constant} := C_1/C_L
    \textcolor{indigo}{local notation} `C' := BET_constant
\end{alltt}

Since \textit{y} and \textit{x} are both functions of pressure, their definitions require pressure as an input. Alternatively, the input can be omitted if we want to deal with x as a function rather than as a number. Notice that the symbols we declared using \textit{constant} do not need to be supplied in the inputs as they already exist in the global workspace. 
 \par We formalize Eq.~\ref{BET 26} by recognizing that the main math behind the BET expression is an infinite sequence that describes the surface area of adsorbed particles for each layer. The series is defined as a function that maps the natural numbers to the real numbers; the natural numbers represent the indexing. It is defined in two cases: if the index is zero, it outputs the surface area of the zeroth layer, and if the index is the \(n+1\), it outputs \(x^{n+1}s_0C\).
\begin{equation}
    \label{BET ith layer}
    s_i = Cx^is_0 \; for \; i : [1,\infty)
\end{equation}
In Lean, we define this sequence as \href{https://atomslab.github.io/LeanChemicalTheories/adsorption/BETInfinite.html#seq}{{\Large\texttwemoji{atom symbol}}}:

\begin{alltt}
    \textcolor{indigo}{def} \textcolor{maroon}{seq} \textcolor{indigo}{(}P : \textcolor{tpurple}{nnreal}\textcolor{indigo}{)} : \textcolor{tpurple}{\(\mathbb{N}\)} → \textcolor{tpurple}{\(\mathbb{R}\)}
    |\textcolor{indigo}{(}0 : \(\mathbb{N}\)\textcolor{indigo}{)}   := s_0
    |\textcolor{indigo}{(}nat.succ n\textcolor{indigo}{)}   := x^\textcolor{indigo}{(}n+1\textcolor{indigo}{)}*s_0*C
\end{alltt}

Where \(s_i\) is the surface area of the i\textsuperscript{th} layer, \(C\) and \(x\) are given by Eq.~\ref{BET C} and Eq.~\ref{BET x}, respectively, and \(s_0\) is the surface area of the zeroth layer. The zeroth layer is the base surface and is constant.

We now have the area and volume equations both in terms of geometric series with well-defined solutions. The BET equation is defined as the ratio of volume absorbed to the volume of a complete unimolecular layer, given by Eq.~\ref{BET Equation}.
\begin{equation}
    \label{BET Equation}
    \frac{V}{A*V_0} = \frac{Cs_0\sum_{i=1} ^{\infty} ix^i}{s_0(1+C\sum_{i=1} ^{\infty} x^i)}
\end{equation}

The main transformation in BET is simplifying this sequence into a simple fraction which involves solving the geometric series. The main math goal is given by Eq.~ \ref{BET Main Math}.
\begin{equation}
    \label{BET Main Math}
    \frac{C\sum_{i=1} ^{\infty} ix^i}{(1+C\sum_{i=1} ^{\infty} x^i)} = \frac{Cx}{(1-x)(1-x+Cx)}
\end{equation}
Before doing the full derivation, we prove Eq.\ref{BET Main Math}, which we call \textit{sequence\_math}. In Lean, this is \href{https://atomslab.github.io/LeanChemicalTheories/adsorption/BETInfinite.html#sequence_math}{{\Large\texttwemoji{atom symbol}}}:
\begin{alltt}
    \textcolor{indigo}{lemma} \textcolor{maroon}{BET.sequence_math} \textcolor{indigo}{\{}P : \textcolor{tpurple}{\(\mathbb{R}\)}\textcolor{indigo}{\}} \textcolor{indigo}{(}hx1: \textcolor{dgreen}{(}x P\textcolor{dgreen}{)} < 1\textcolor{indigo}{)} \textcolor{indigo}{(}hx2 : 0 < \textcolor{dgreen}{(}x P\textcolor{dgreen}{)}\textcolor{indigo}{)} :
        \textcolor{indigo}{(}\(\sum\)' k : \(\mathbb{N}\), \textcolor{dgreen}{(}\textcolor{iibrown}{(}k + 1\textcolor{iibrown}{)}*\textcolor{iibrown}{(}\textcolor{maroon}{seq} P \textcolor{dorange}{(}k+1\textcolor{dorange}{)}\textcolor{iibrown}{)}\textcolor{dgreen}{)}\textcolor{indigo}{)}/\textcolor{indigo}{(}s_0 + \(\sum\)' k, \textcolor{dgreen}{(}\textcolor{maroon}{seq} P \textcolor{iibrown}{(}k+1\textcolor{iibrown}{)}\textcolor{dgreen}{)}\textcolor{indigo}{)} = 
        C*\textcolor{indigo}{(}x P\textcolor{indigo}{)}/\textcolor{indigo}{(}\textcolor{dgreen}{(}1 -  \textcolor{iibrown}{(}x P\textcolor{iibrown}{)}\textcolor{dgreen}{)}*\textcolor{dgreen}{(}1 - \textcolor{iibrown}{(}x P\textcolor{iibrown}{)} + \textcolor{iibrown}{(}x P\textcolor{iibrown}{)}*C\textcolor{dgreen}{)}\textcolor{indigo}{)} :=
\end{alltt}

In Lean, the apostrophe after the sum symbol denotes an infinite sum, which is defined to start at zero since it is indexed by the natural numbers, which start at zero. Since the infinite sum of Eq.~\ref{BET Main Math} starts at one, we add one to all the indexes, \textit{k}, so that when $k$ is zero, we get one, etc. We also define two new theorems that derive the solution to these geometric series with an index starting at one. After expanding \textit{seq}, we use those two theorems, and then rearrange the goal to get two sides that are equal. We also use the tag \textit{lemma} instead of \textit{theorem}, just to communicate that it is a lower-priority theorem, intended to prove other theorems. The tag \textit{lemma} has no functional difference from \textit{theorem} in Lean, it's purpose is for mathematicians to label proofs. 
\par With this we can formalize the derivation of Eq.~\ref{BET 26}. First we define Eq.~\ref{BET 26} as a new object and then prove a theorem showing we can derive this object from the sequence. In Lean, the definition looks like this \href{https://atomslab.github.io/LeanChemicalTheories/adsorption/BETInfinite.html#brunauer_26}{{\Large\texttwemoji{atom symbol}}}:

\begin{alltt}
    \textcolor{indigo}{def} \textcolor{maroon}{brunauer_26} := \(\lambda\) P :  \textcolor{indigo}{\textcolor{dgreen}{\textcolor{tpurple}{\(\mathbb{R}\)}}}, C* \textcolor{indigo}{(}x P\textcolor{indigo}{)}/\textcolor{indigo}{(}\textcolor{dgreen}{(}1-\textcolor{iibrown}{(}x P\textcolor{iibrown}{)}\textcolor{dgreen}{)}*\textcolor{dgreen}{(}1-\textcolor{iibrown}{(}x P\textcolor{iibrown}{)}+C*\textcolor{iibrown}{(}x P\textcolor{iibrown}{)}\textcolor{dgreen}{)}\textcolor{indigo}{)}
\end{alltt}

Here, we explicitly define this as a function, because we want to deal with Eq.~\ref{BET 26} normally as a function of pressure, rather then just a number. Now we can prove a theorem that formalizes the derivation of this equation \href{https://atomslab.github.io/LeanChemicalTheories/adsorption/BETInfinite.html#brunauer_26_from_seq}{{\Large\texttwemoji{atom symbol}}}:

\begin{alltt}
    \textcolor{indigo}{theorem} \textcolor{maroon}{brunauer_26_from_seq}
    \textcolor{indigo}{\{}P V_0 : \textcolor{dgreen}{\textcolor{tpurple}{\(\mathbb{R}\)}}\textcolor{indigo}{\}}
    (hx1: (x P) < 1)
    \textcolor{indigo}{(}hx2 : 0 < \textcolor{dgreen}{(}x P\textcolor{dgreen}{)}\textcolor{indigo}{)}
    :
      \textcolor{indigo}{let} Vads :=  V_0 * \(\sum'\) \textcolor{indigo}{(}k : \(\mathbb{N}\)\textcolor{indigo}{)}, ↑k * \textcolor{indigo}{(}seq P k\textcolor{indigo}{)},
          A :=  \(\sum'\) \textcolor{indigo}{(}k : \(\mathbb{N}\)), \textcolor{indigo}{(}seq P k\textcolor{indigo}{)} \textcolor{indigo}{in}
      Vads/A = V_0*\textcolor{indigo}{(}\textcolor{maroon}{brunauer_26} P\textcolor{indigo}{)}
    :=
\end{alltt}

Unlike the Langmuir proof introduced earlier in Fig.~\ref{LangmuirProof}, the BET uses definitions that allow reusability of those definitions across the proof structure. The proof starts by showing that \textit{seq} is summable. This means the sequence has some infinite sum and the \(\sum'\) symbol is used to get the value of that infinite series. We show in the proof that both \textit{seq} and \textit{k*seq} is summable, where the first is needed for the area sum and the second is needed for the volume sum. After that, we simplify our definitions, move the index of the sum from zero to one so we can simplify the sequence, and apply the \textit{BET.sequence\_math} lemma we proved above. Finally, we use the \textit{field\_simp} tactic to rearrange and close the goal. With that, we formalized the derivation of Eq.~\ref{BET 26}, just as Brunauer, \emph{et al.} did in 1938.
\par In the SI, we continue formalizing BET theory by deriving Equation 28 from the paper, given by Eq.~\ref{BET 28.2}
\begin{equation}
    \label{BET 28.2}
     \frac{V}{A*V_0} = \frac{CP}{(P_0-P)(1+(C-1)(P/P_0))}
\end{equation}
This follows from recognizing that $1/C_L = P_0$. While Brunauer, \emph{et al.} attempt to show this in the paper, we discuss the trouble with implementing the logic they present. Instead, we show a similar proof that Eq.~\ref{BET 26} approaches infinity as pressure approaches $1/C_L$, and assume as a premise in the derivation of Eq.~\ref{BET 28.2} that $1/C_L  \equiv P_0$.

\subsection{Classical Thermodynamics and Gas Laws: Introducing Lean Structures}

Lean is so expressive because it enables relationships between mathematical objects. We can use this functionality to precisely define and relate \emph{scientific concepts} with mathematical certainty. We illustrate this by formalizing proofs of gas laws in classical thermodynamics.

\begin{figure}[h]
\centering
\BeginAccSupp{ActualText={This is a description of the 'thermo_system' and 'ideal_gas' structures in the Lean theorem prover. These structures represent different types of thermodynamic systems, such as 'isobaric', 'isochoric', and 'isothermal'. They use Lean's definitions to help in proving theorems related to gas laws.}}
\includegraphics[width=\textwidth]{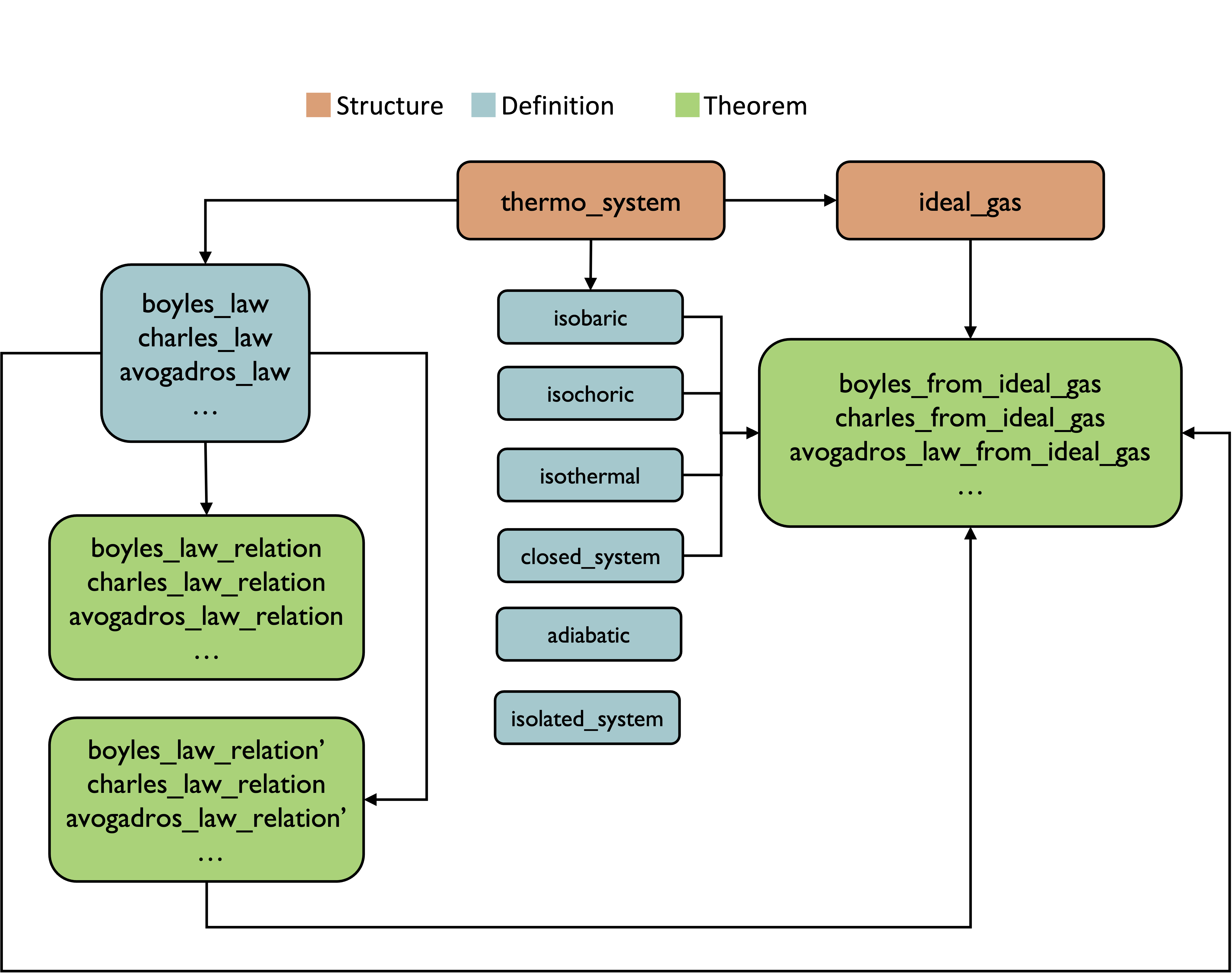}
\EndAccSupp{}
\caption{Thermodynamic system in Lean. Here the \textit{thermo\_system} and \textit{ideal\_gas} are Lean \texttt{structures} that describe different kinds of thermodynamic systems like \textit{isobaric}, \textit{isochoric},  \textit{isothermal} etc. using Lean \texttt{definitions} to proof \texttt{theorems} relating to the \textit{gas laws}. }
\label{thermo}
\end{figure}

We can prove that the ideal gas law, $PV=nRT$ follows Boyle's Law, $P_1 V_1 = P_2 V_2$, following the style of our derivation of Langmuir's theory: demonstrating that a conjecture follows from the premises \href{https://atomslab.github.io/LeanChemicalTheories/thermodynamics/boyles_law.html}{{\Large\texttwemoji{atom symbol}}}. However, this proof style doesn't facilitate interoperability among proofs and limits the mathematics that can be expressed.
z
In contrast, we can prove the same, more systematically, by first formalizing the concepts of thermodynamic systems and states, extending that system to a specific ideal gas system, defining Boyle's Law in light of these thermodynamic states, and then proving that the ideal gas obeys Boyle's Law (see Fig.~\ref{thermo}).

Classical thermodynamics describes the macroscopic properties of thermodynamic states and relationships between them \cite{dahm_visco_2015, sandler2017chemical}. We formalize the concept of ``thermodynamic system'' by defining a Lean \texttt{structure} called \textit{thermo\_system} over the real numbers, with thermodynamic properties (e.g., pressure, volume, etc.) defined as functions from a type to the real numbers \(\alpha \rightarrow \mathbb{R}\). Here, \(\alpha\) is meant to represent a general indexing type. It could be the natural numbers if we wanted to use those to represent states of the system, real numbers to represent time, or anything else. The only requirement is that \(\alpha\) is nontrivial, meaning it has at least two different elements. In Lean, this is \href{https://atomslab.github.io/LeanChemicalTheories/thermodynamics/basic.html#thermo_system}{{\Large\texttwemoji{atom symbol}}}:

\begin{alltt}
    \textcolor{indigo}{structure} \textcolor{maroon}{thermo_system}  \textcolor{indigo}{(}\textcolor{tpurple}{\(\alpha\)}\textcolor{indigo}{)}  \textcolor{indigo}{[}nontrivial \textcolor{tpurple}{\(\alpha\)}\textcolor{indigo}{]}:=
    \textcolor{indigo}{(}pressure : \textcolor{tpurple}{\(\alpha\)} → \textcolor{tpurple}{\(\mathbb{R}\)}\textcolor{indigo}{)}
    \textcolor{indigo}{(}volume :  : \textcolor{tpurple}{\(\alpha\)} → \textcolor{tpurple}{\(\mathbb{R}\)}\textcolor{indigo}{)}
    \textcolor{indigo}{(}temperature : \textcolor{tpurple}{\(\alpha\)} → \textcolor{tpurple}{\(\mathbb{R}\)}\textcolor{indigo}{)}
    \textcolor{indigo}{(}substance_amount : \textcolor{tpurple}{\(\alpha\)} → \textcolor{tpurple}{\(\mathbb{R}\)}\textcolor{indigo}{)}
    \textcolor{indigo}{(}energy : \textcolor{tpurple}{\(\alpha\)} → \textcolor{tpurple}{\(\mathbb{R}\)}\textcolor{indigo}{)}
\end{alltt}

We define six descriptions of the system: isobaric (constant pressure); isochoric (constant volume); isothermal (constant temperature); adiabatic (constant energy); closed (constant mass); and isolated (constant mass and energy). Each of these conditions has the type \textit{Prop}, or proposition, considering them to be assertions about the system. We formally define these by stating that, for all ($\forall$) pairs of states $n$ and $m$, the property at those states is equal. We define these six descriptions to take in a \textit{thermo\_system} since we need to specify what system we are ascribing this property to.  In Lean, this is \href{hhttps://atomslab.github.io/LeanChemicalTheories/thermodynamics/basic.html#isobaric}{{\Large\texttwemoji{atom symbol}}}:

\begin{alltt}
    \textcolor{indigo}{def} \textcolor{maroon}{isobaric} \textcolor{indigo}{(}\textcolor{tpurple}{\(\alpha\)}\textcolor{indigo}{)} \textcolor{indigo}{[}nontrivial \textcolor{tpurple}{\(\alpha\)}\textcolor{indigo}{]} \textcolor{indigo}{(}M : \textcolor{maroon}{thermo_system} \textcolor{indigo}{)} : \textcolor{tpurple}{Prop} := 
    \(\forall\) n m : \textcolor{tpurple}{\(\alpha\)}, pressure n = pressure m
    \textcolor{indigo}{def} \textcolor{maroon}{isochoric} \textcolor{indigo}{(}\textcolor{tpurple}{\(\alpha\)}\textcolor{indigo}{)} \textcolor{indigo}{[}nontrivial \textcolor{tpurple}{\(\alpha\)}\textcolor{indigo}{]} \textcolor{indigo}{(}M : \textcolor{maroon}{thermo_system} \textcolor{indigo}{)} : \textcolor{tpurple}{Prop} :=  
    \(\forall\) n m : \textcolor{tpurple}{\(\alpha\)}, volume n = volume m
    \textcolor{indigo}{def} \textcolor{maroon}{isothermal} \textcolor{indigo}{(}\textcolor{tpurple}{\(\alpha\)}\textcolor{indigo}{)} \textcolor{indigo}{[}nontrivial \textcolor{tpurple}{\(\alpha\)}\textcolor{indigo}{]} \textcolor{indigo}{(}M : \textcolor{maroon}{thermo_system} \textcolor{indigo}{)} : \textcolor{tpurple}{Prop}:= 
    \(\forall\) n m : \textcolor{tpurple}{\(\alpha\)}, temperature n = temperature m
    \textcolor{indigo}{def} \textcolor{maroon}{adiabatic} \textcolor{indigo}{(}\textcolor{tpurple}{\(\alpha\)}\textcolor{indigo}{)} \textcolor{indigo}{[}nontrivial \textcolor{tpurple}{\(\alpha\)}\textcolor{indigo}{]} \textcolor{indigo}{(}M : \textcolor{maroon}{thermo_system} \textcolor{indigo}{)} : \textcolor{tpurple}{Prop }:= 
    \(\forall\) n m : \textcolor{tpurple}{\(\alpha\)}, energy n = energy m
    \textcolor{indigo}{def} \textcolor{maroon}{closed_system}  \textcolor{indigo}{(}\textcolor{tpurple}{\(\alpha\)}\textcolor{indigo}{)} \textcolor{indigo}{[}nontrivial \textcolor{tpurple}{\(\alpha\)}\textcolor{indigo}{]} \textcolor{indigo}{(}M : \textcolor{maroon}{thermo_system} \textcolor{indigo}{)} : \textcolor{tpurple}{Prop}:= 
    \(\forall\) n m : \textcolor{tpurple}{\(\alpha\)}, substance_amount n = substance_amount m
    \textcolor{indigo}{def} \textcolor{maroon}{isolated_system}  \textcolor{indigo}{(}\textcolor{tpurple}{\(\alpha\)}\textcolor{indigo}{)} \textcolor{indigo}{[}nontrivial \textcolor{tpurple}{\(\alpha\)}\textcolor{indigo}{]} \textcolor{indigo}{(}M : \textcolor{maroon}{thermo_system} \textcolor{indigo}{)} : \textcolor{tpurple}{Prop} := 
    adiabatic M \(\wedge\) closed_system M
\end{alltt}

We define an isolated system as just a closed system and ($\wedge$) adiabatic, rather than using the universal quantifier ($\forall$), since it would be redundant.
\par Now that the basics of a thermodynamic system have been defined, we can define models that attempt to describe the system mathematically. These models can be defined as another structure, which \textit{extends} the \textit{thermo\_system} structure. When a structure extends another structure, it inherits the properties of the structure it extended. This allows us to create a hierarchy of structures so we don't have to redefine properties repeatedly. The most well-known model is the ideal gas model, which comes with the ideal gas law equation of state. We define the ideal gas model to have two properties, the universal gas constant, \(R\), and the ideal gas law. In the future, we plan to add more properties to the definition, especially as we expand on the idea of energy. We define the ideal gas law as an equation relating the products of pressure and volume to the product of temperature, amount of substance, and the gas constant. In Lean, this is \href{https://atomslab.github.io/LeanChemicalTheories/thermodynamics/basic.html#ideal_gas}{{\Large\texttwemoji{atom symbol}}}:

\begin{alltt}
    \textcolor{indigo}{structure} \textcolor{maroon}{ideal_gas}
      \textcolor{indigo}{extends} \textcolor{maroon}{thermo_system} :=
    \textcolor{indigo}{(}R : \textcolor{tpurple}{\(\mathbb{R}\)}\textcolor{indigo}{)}
    \textcolor{indigo}{(}ideal_gas_law : \(\forall\) n : \textcolor{tpurple}{\(\alpha\)},\textcolor{dgreen}{(}pressure n\textcolor{dgreen}{)}*\textcolor{dgreen}{(}volume n\textcolor{dgreen}{)} = \textcolor{dgreen}{(}substance_amount n\textcolor{dgreen}{)}*R*\textcolor{dgreen}{(}temperature n\textcolor{dgreen}{)}\textcolor{indigo}{)}
\end{alltt}

To define a system modeled as an ideal gas, we write in Lean: \textit{(M : ideal\_gas $\mathbb{R}$)}. Now we have a system, M, modeled as an ideal gas. 
\par Boyle's law states that the pressure of an ideal gas is inversely proportional to the system's volume in an isothermal and closed system \cite{levine_1978}. This is mathematically given by Eq.~\ref{Boyle's Law}, where \(P\) is pressure, \(V\) is volume, and \(k\) is a constant whose value is dependent on the system. 

\begin{equation}
    \label{Boyle's Law}
    PV = k
\end{equation}

In Lean, we define Boyle's Law as \href{https://atomslab.github.io/LeanChemicalTheories/thermodynamics/basic.html#boyles_law}{{\Large\texttwemoji{atom symbol}}}:

\begin{alltt}
    \textcolor{indigo}{def} \textcolor{maroon}{boyles_law}  \textcolor{indigo}{\{}\textcolor{tpurple}{\(\alpha\)}\textcolor{indigo}{\}} \textcolor{indigo}{[}nontrivial \textcolor{tpurple}{\(\alpha\)}\textcolor{indigo}{]} \textcolor{indigo}{(}M : \textcolor{maroon}{thermo_system \textcolor{tpurple}{\(\alpha\)}} \textcolor{indigo}{)} := 
    \(\exists\)\textcolor{indigo}{(}k : \textcolor{tpurple}{\(\mathbb{R}\)}\textcolor{indigo}{)}, \(\forall\) n : \textcolor{tpurple}{\(\alpha\)}, \textcolor{indigo}{(}pressure n\textcolor{indigo}{)} * \textcolor{indigo}{(}volume n\textcolor{indigo}{)}= k
\end{alltt}

We use the existential operator ($\exists$) on \(k\), which can be read as \textit{there exists a \(k\)}, because each system has a specific constant. We also define the existential before the universal, so it is logically correct. Right now, it reads, \textit{there exists a \(k\), such that for all states, this relation holds}. If we write it the other way, it would say \textit{for all states, there exists a \(k\), such that this relation holds.} The second way means that \(k\) is dependent on the state of the system, which isn't true. The constant is the same for any state of a system. Also, even though Boyle's law is a statement about an ideal gas, we define it as a general system so, in the future, we can look at what assumptions are needed for other models to obey Boyle's Law. 
\par Next, we prove a couple of theorems relating to the relations that can be derived from Boyle's law. From Eq.~\ref{Boyle's Law}, we can derive a relation between any two states, given by Eq.~\ref{Boyle's Relation}, where \(n\) and \(m\) are two states of the system.

\begin{equation}
    \label{Boyle's Relation}
    P_nV_n = P_mV_m
\end{equation}

The first theorem we prove shows how Eq.\ref{Boyle's Relation} follows from Eq.\ref{Boyle's Law}. In Lean this looks like \href{https://atomslab.github.io/LeanChemicalTheories/thermodynamics/basic.html#boyles_law_relation}{{\Large\texttwemoji{atom symbol}}}:

\begin{alltt}
    \textcolor{indigo}{theorem} \textcolor{maroon}{boyles_law_relation} \textcolor{indigo}{\{}\textcolor{tpurple}{\(\alpha\)}\textcolor{indigo}{\}} \textcolor{indigo}{[}nontrivial \textcolor{tpurple}{\(\alpha\)}\textcolor{indigo}{]} \textcolor{indigo}{(}M : \textcolor{maroon}{thermo_system \textcolor{tpurple}{\(\alpha\)}}\textcolor{indigo}{)} : 
    \textcolor{maroon}{boyles_law} M → \(\forall\) n m : \textcolor{tpurple}{\(\alpha\)}, pressure n * volume n = pressure m * volume m :=
\end{alltt}

The right arrow can be read as \textit{implies}, so the statement says that \textit{Boyle's law implies Boyle's relation}. This is achieved using modus ponens, introducing two new names for the universal quantifier, then rewriting Boyle's law into the goal by specializing Boyle's law with \(n\) and \(m\). We also want to show that the inverse relation holds, such that Eq.~\ref{Boyle's Relation} implies Eq.~\ref{Boyle's Law}. In Lean, this is \href{https://atomslab.github.io/LeanChemicalTheories/thermodynamics/basic.html#boyles_law_relation'}{{\Large\texttwemoji{atom symbol}}}:

\begin{alltt}
    \textcolor{indigo}{theorem} \textcolor{maroon}{boyles_law_relation'} \textcolor{indigo}{\{}\textcolor{tpurple}{\(\alpha\)}\textcolor{indigo}{\}} \textcolor{indigo}{[}nontrivial \textcolor{tpurple}{\(\alpha\)}\textcolor{indigo}{]} \textcolor{indigo}{(}M : \textcolor{maroon}{thermo_system \textcolor{tpurple}{\(\alpha\)}}\textcolor{indigo}{)} : 
    \textcolor{indigo}{(}\(\forall\) n m, pressure n * volume n = pressure m * volume m\textcolor{indigo}{)} → \textcolor{maroon}{boyles_law} M :=
\end{alltt}

We begin in the same way by using modus ponens and simplifying Boyle's law in the form of Eq.~\ref{Boyle's Law}. Next, we satisfy the existential by providing an old name. In our proof, we use \(P_1V_1\) as an old name for \(k\), then we specialize the relation with \(n\) and \(1\) and close the goal. 
\par Finally, with these two theorems, we show that Boyle's law can be derived from the ideal gas law under the assumption of an isothermal and closed system. In Lean, this is \href{https://atomslab.github.io/LeanChemicalTheories/thermodynamics/basic.html#boyles_from_ideal_gas}{{\Large\texttwemoji{atom symbol}}}:

\begin{alltt}
    \textcolor{indigo}{theorem} \textcolor{maroon}{boyles_from_ideal_gas} \textcolor{indigo}{\{}\textcolor{tpurple}{\(\alpha\)}\textcolor{indigo}{\}} \textcolor{indigo}{[}nontrivial \textcolor{tpurple}{\(\alpha\)}\textcolor{indigo}{]} \textcolor{indigo}{(}M : \textcolor{maroon}{ideal_gas \textcolor{tpurple}{\(\alpha\)}}\textcolor{indigo}{)}  
     \textcolor{indigo}{(}iso1 : \textcolor{maroon}{isothermal} M.to_\textcolor{maroon}{thermo_system} \textcolor{indigo}{)} \textcolor{indigo}{(}iso2 : \textcolor{maroon}{closed_system} M.to_\textcolor{maroon}{thermo_system} \textcolor{indigo}{)}: 
    \textcolor{maroon}{boyles_law} M :=
\end{alltt}

This proof is completed by using the second theorem for Boyle's relation and simplifying the ideal gas relation using the two \textit{iso} constraints. 

We have implemented this framework to prove both Charles' and Avagadro's Law \href{https://atomslab.github.io/LeanChemicalTheories/thermodynamics/basic.html}{{\Large\texttwemoji{atom symbol}}} illustrating the interoperability of these proofs. In the future, we plan to define energy and prove theorems relating to it, including the laws of thermodynamics \cite{atkins2010laws}.

\subsection{Kinematic equations: Calculus in Lean}
Calculus and differential equations are ubiquitous in chemical theory, and much has been formalized in \texttt{mathlib}. To illustrate Lean's calculus capabilities and motivate future formalization efforts, we formally prove that the kinematic equations follow from calculus-based definitions of motion, assuming constant acceleration. The analysis of physical equations of motion, particularly those based on Newtonian mechanics, is strongly related to the formulation of many theories in chemical physics, including reaction kinetics \cite{frost1961kinetics}, diffusion and transport phenomena \cite{bird2002transport} and molecular dynamics \cite{haile1993molecular}. These concepts are essential for understanding chemical reactions and how molecules move and interact. 

%Equations of motion are the basis for many chemical theories, such as reaction kinetics \cite{frost1961kinetics} and molecular dynamics \cite{haile1993molecular} that use Newtonian mechanics. 

\par The equations of motion are a set of two coupled differential equations that relate the position, velocity, and acceleration of an object in an n-dimensional vector space \cite{beggs1983kinematics}. The differential equations are given by Eq.~\ref{Differential Equation 1} and \ref{Differential Equation 2}, where \textbf{x}, \textbf{v}, and \textbf{a} represent position, velocity, and acceleration, respectively (bold type face signifies a vector quantity). All three variables are parametric equations, where each dimension of the vector is a function of time.\footnote{These proofs could also be constructed using partial differential equations, but \texttt{mathlib} doesn't currently have enough theorems for partial derivatives.}

\begin{equation}
    \label{Differential Equation 1}
    \textbf{v}(t) = \frac{d(\textbf{x}(t))}{dt}
\end{equation}
\begin{equation}
    \label{Differential Equation 2}
    \textbf{a}(t) = \frac{d(\textbf{v}(t))}{dt}
\end{equation}

\begin{figure}[h]
\centering
\BeginAccSupp{ActualText={This is a description of how kinematics is represented in the Lean theorem prover. A structure named 'motion' is defined to depict the relationship between position, velocity, and acceleration through the use of differential equations. These relationships are formalized and proven using Lean's definitions of derivative functions.}}
\includegraphics[width=0.8\textwidth]{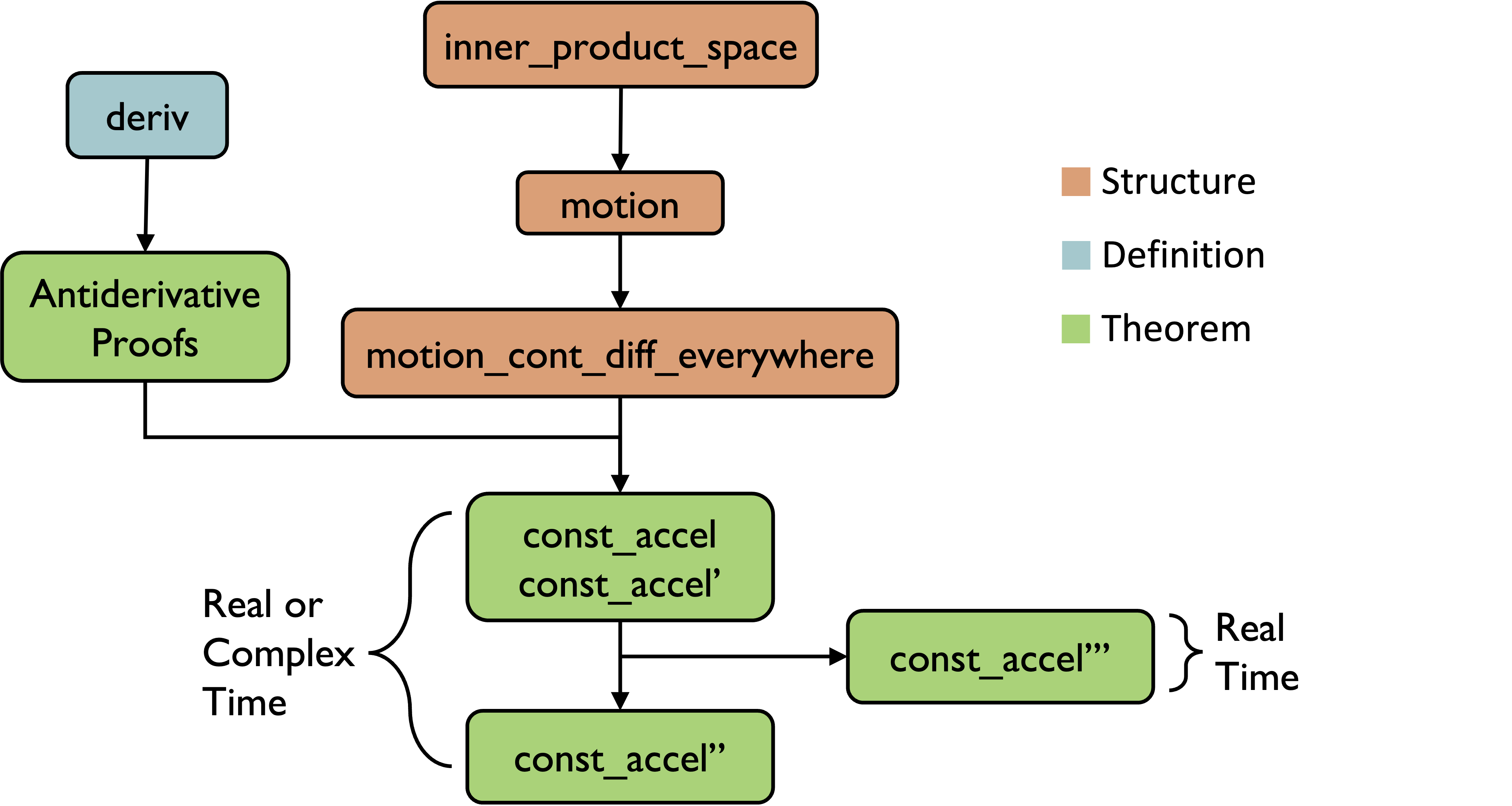}
\caption{Kinematics in Lean. Here we define \textit{motion} as Lean \texttt{structure} that represents the relation between position, velocity, and acceleration through differential equations that are proved using \texttt{definitions} of \textit{derivative} functions.}
\EndAccSupp{}
\label{kinematicsFig}
\end{figure}

As in the thermodynamics section, we can define a structure, \texttt{motion}, to encompass these concepts. This structure defines three new elements: position, velocity, and acceleration, which are functions, and two differential equations relating these three functions. This structure also requires the vector space to form an inner product space, which is a real or complex vector space with an operator (the inner product) over the field. The inner product is a generalization of the dot product for any vector space. By requiring \texttt{inner\_product\_space}, the motion structure inherits all of \texttt{inner\_product\_space}'s properties and allows us to access the calculus theorems in \texttt{mathlib}. In Lean, this is \href{https://atomslab.github.io/LeanChemicalTheories/physics/kinematic_equations.html#motion}{{\Large\texttwemoji{atom symbol}}}:

\begin{alltt}
     \textcolor{indigo}{structure}  \textcolor{maroon}{motion} \textcolor{indigo}{(}\textcolor{tpurple}{\(\mathbb{K}\)} : Type u_1\textcolor{indigo}{)} \textcolor{indigo}{(}\textcolor{tpurple}{E} : Type u_2\textcolor{indigo}{)} \textcolor{indigo}{[}is_R_or_C \textcolor{tpurple}{\(\mathbb{K}\)}\textcolor{indigo}{]}
     \textcolor{indigo}{[}inner_product_space \textcolor{tpurple}{\(\mathbb{K}\)} \textcolor{tpurple}{\(\mathbb{E}\)}\textcolor{indigo}{]} :=
    \textcolor{indigo}{(}position velocity acceleration : \textcolor{tpurple}{\(\mathbb{K}\)} → \textcolor{tpurple}{E} \textcolor{indigo}{)} 
    \textcolor{indigo}{(}hvel : velocity = deriv position\textcolor{indigo}{)}
    \textcolor{indigo}{(}hacc : acceleration = deriv velocity\textcolor{indigo}{)}
\end{alltt}
$\mathbb{K}$ represents a field that we require to be either a real ($\mathbb{R}$) or complex ($\mathbb{C}$) number, and \(E\) symbolizes a general vector field. In mathematics, a field is an algebraic structure with addition, subtraction, multiplication, and division operations. Our vector space could be an n-dimensional Euclidean vector space, but we instead use a general vector field to be as general as possible. This allows us to describe motion in a Euclidean vector space, as well as a hyperbolic vector space or a vector space with special properties. 

\par In Lean, if a function is not differentiable at a point, the derivative at that point returns zero\footnote{Likewise, division by zero is defined to return zero instead of something like "undefined" or "NaN”. In Lean and other theorem provers, the symbol / is not used for mathematical division but instead points to a function called \emph{real.div}. This function returns $x/y$ if $y$ is not zero, and 0 if $y$ equals 0. Another case is the real square root function (\emph{real.sqrt}), which outputs a real number for any input, even negatives since it is defined as $\mathbb{R}$ → $\mathbb{R}$. These conventions may be unfamiliar to scientists and engineers, but they are used for convenience and won't lead to contradictions in a proof. Any invalid step in a proof involving these conventions will be caught when invoking a theorem not true for its definition. We wrestled with this convention for some time before finding clarity  \href{https://web.archive.org/web/20230719150030/https://xenaproject.wordpress.com/2020/07/05/division-by-zero-in-type-theory-a-faq/}{\color{blue}{in this blog post}} archived for reference \cite{archiveDivisionZero}.}. During our first formalization attempt, we tried to define a function to be constant by setting its derivative to zero. However, $df/dx = 0$ may also arise if a function is not differentiable at that point. To avoid this edge case, we define another structure to require the equations of motion to be n-times continuously differentiable everywhere. We only require the equations to be n-times differentiable instead of infinitely differentiable for generality reasons, however, a theorem can instantiate this structure and assume infinite differentiability. We also declare this as a separate structure, instead of in the \textit{motion} structure, to allow future proofs that require the equations to be n-times continuously differentiable on a set or an interval rather than everywhere (e.g., a molecular mechanics force field with a non-smoothed cutoff is not differentiable at that point). That way, depending on the theorem, the user can choose the appropriate extension. In Lean, this structure looks like \href{https://atomslab.github.io/LeanChemicalTheories/physics/kinematic_equations.html#motion_cont_diff_everywhere}{{\Large\texttwemoji{atom symbol}}}:
\begin{alltt}
    \textcolor{indigo}{structure}  \textcolor{maroon}{motion_cont_diff_everywhere} \textcolor{indigo}{(}\textcolor{tpurple}{\(\mathbb{K}\)} :Type u_1\textcolor{indigo}{)} \textcolor{indigo}{(}\textcolor{tpurple}{E} : Type u_2\textcolor{indigo}{)} \textcolor{indigo}{[}is_R_or_C \textcolor{tpurple}{\(\mathbb{K}\)}\textcolor{indigo}{]} 
    \textcolor{indigo}{[}inner_product_space \textcolor{tpurple}{\(\mathbb{K}\)} \textcolor{tpurple}{\(\mathbb{E}\)}\textcolor{indigo}{]}
      \textcolor{indigo}{extends} \textcolor{maroon}{motion} \textcolor{tpurple}{\(\mathbb{K}\)} \textcolor{tpurple}{E} :=
    \textcolor{indigo}{(}contdiff : \(\forall\) n : with_top \textcolor{tpurple}{\(\mathbb{N}\)}, ∀ m : \textcolor{tpurple}{\(\mathbb{N}\)}, 
    \textcolor{dgreen}{(}m < n\textcolor{dgreen}{)} \(\rightarrow\) \textcolor{dgreen}{(}cont_diff \textcolor{tpurple}{\(\mathbb{K}\)} n \textcolor{iibrown}{(}deriv^\textcolor{indigo}{[}m\textcolor{indigo}{]} position\textcolor{iibrown}{)}\textcolor{dgreen}{)}\textcolor{indigo}{)}
\end{alltt}
The field \textit{contdiff} states that for all $n$, defined as a natural number including positive infinity, and for all $m$, defined as a natural number, if $m$ is less than $n$, then the \(m^{th}\) derivative of position is continuously differentiable $n$-times. 

\par When acceleration is constant, %, meaning the acceleration points linearly in one direction,
this set of differential equations has four useful analytical solutions, the kinematic equations, Eq.~\ref{Kinematic Equation 1}--\ref{Kinematic Equation 4}, where the subscript naught denotes variables evaluated at $t=0$. 
\begin{equation}
    \label{Kinematic Equation 1}
    \textbf{v}(t) = \textbf{a}t+\textbf{v}_0
\end{equation}
\begin{equation}
    \label{Kinematic Equation 2}
    \textbf{x}(t) = \frac{\textbf{a}t^2}{2}+\textbf{v}_0t+\textbf{x}_0
\end{equation}
\begin{equation}
\label{Kinematic Equation 3}
    \textbf{x}(t) = \frac{\textbf{v}(t)+\textbf{v}_0}{2}t+\textbf{x}_0
\end{equation}
\begin{equation}
    \label{Kinematic Equation 4}
    v^2(t) = v_0^2+2\textbf{a}\cdot \textbf{d}
\end{equation}

Under the assumption of one-dimensional motion, these equations simplify to the familiar introductory kinematic equations. Eq.~\ref{Kinematic Equation 4}, also known as the Torricelli Equation, uses the shorthand square to represent the dot product, \(v^2(t) \equiv \textbf{v(t)}\cdot \textbf{v(t)}\). 
\par With this, we can now begin deriving the four kinematic equations. The first three derivations for Eq.~\ref{Kinematic Equation 1}--\ref{Kinematic Equation 3}, all use the same premises, given below:
\begin{alltt}
    \textcolor{indigo}{(}\textcolor{tpurple}{\(\mathbb{K}\)} : Type u_1\textcolor{indigo}{)} \textcolor{indigo}{(}\textcolor{tpurple}{E} : Type u_2\textcolor{indigo}{)} [is_R_or_C \textcolor{tpurple}{\(\mathbb{K}\)}] [inner_product_space \textcolor{tpurple}{\(\mathbb{K}\)} E] 
    \textcolor{indigo}{(}M : motion_cont_diff_everywhere \textcolor{tpurple}{\(\mathbb{K}\)} \textcolor{tpurple}{E}\textcolor{indigo}{)} 
    \textcolor{indigo}{(}A : \textcolor{tpurple}{E}) \textcolor{indigo}{(}n : with_top \textcolor{tpurple}{\(\mathbb{N}\)}\textcolor{indigo}{)} 
    \textcolor{indigo}{(}accel_const : motion.acceleration = \(\lambda\) \textcolor{dgreen}{(}t : \textcolor{tpurple}{\(\mathbb{K}\)}\textcolor{dgreen}{)}, A\textcolor{indigo}{)} 
\end{alltt}
The first line contains four premises to declare the field and vector space the motion space is defined on. The next line defines a motion space, M. The third line contains two premises, a variable, A, which represents the value of constant acceleration, and n, the number of times position can be differentiated. When applying these theorems, the \textit{top} function, which means positive infinity in Lean, can be used to specify n. The final line is a premise that assumes acceleration is constant. The lambda function is constant because A is not a function of t, so for any value of t, the function outputs the same value, A. The three kinematic equations in Lean \href{https://atomslab.github.io/LeanChemicalTheories/physics/kinematic_equations.html#const_accel}{{\Large\texttwemoji{atom symbol}}} are given below (note, the premises are omitted since they have already been given above). 
\begin{alltt}
    \textcolor{indigo}{theorem} \textcolor{maroon}{const_accel} premises : velocity = \(\lambda\) \textcolor{indigo}{(}t : \textcolor{tpurple}{\(\mathbb{K}\)}\textcolor{indigo}{)}, t\(\cdot\)\textbf{A} + velocity 0 :=
    
    \textcolor{indigo}{theorem} \textcolor{maroon}{const_accel'} premises :
    position = \(\lambda\) \textcolor{indigo}{(}t : \textcolor{tpurple}{\(\mathbb{K}\)}\textcolor{indigo}{)}, \textcolor{indigo}{(}t^2\textcolor{indigo}{)}/2\(\cdot\)A + t\(\cdot\)\textcolor{indigo}{(}velocity 0\textcolor{indigo}{)} + position 0 :=
    
    \textcolor{indigo}{theorem} \textcolor{maroon}{const_accel''} premises :
    \(\forall\) t : \textcolor{tpurple}{\(\mathbb{K}\)}, position t = \textcolor{indigo}{(}t/2\textcolor{indigo}{)}\(\cdot\)\textcolor{indigo}{(}\textcolor{dgreen}{(}velocity t\textcolor{dgreen}{)} - \textcolor{dgreen}{(}velocity 0\textcolor{dgreen}{)}\textcolor{indigo}{)} + position 0 := 
\end{alltt}
The \textit{\(\cdot\)} symbol indicates scalar multiplication, such as when a vector is multiplied by a scalar. We normally use the \textit{\(\cdot\)} symbol for the dot product, but Lean uses the \texttt{inner} function for the dot product. Also, \texttt{velocity 0} means the velocity function evaluated at 0. Lean uses parentheses for orders of operations, not for function inputs, so \(f(x)\) in normal notation converts to \texttt{f x} in Lean. The proofs of the first two theorems use the two differential equations from the \texttt{motion} structure and the antiderivative, whose formalization we explain in the supplementary information (these theorems weren't available in \texttt{mathlib} at the time of writing, so we proved them ourselves). The third theorem is proved by rearranging the previous two theorems.

Because we declared the field \texttt{is\_R\_or\_C}, the above proofs hold for both real and complex time. However, we were unable to prove Eq.~\ref{Kinematic Equation 4}, due to the complex conjugate that arises when simplifying the proof. Eq.~\ref{Kinematic Equation 4} uses the inner product, a function that takes in two vectors from a vector space and outputs a scalar. If the vector space is a Euclidean vector space, this is just the dot product. The inner product is semi-linear, linear in its first argument, Equation \ref{Inner Product Linear}, but sesquilinear in its second argument, Equation \ref{Inner Product Sesquilinear}.
\begin{gather}
    \label{Inner Product Linear}
    \langle ax+by, z \rangle = a \langle x,z \rangle + b \langle y,z \rangle
    \\
    \label{Inner Product Sesquilinear}
    \langle x,ay + bz \rangle = \Bar{a}\langle x,y \rangle + \Bar{b}\langle x,z \rangle
\end{gather}
The bar denotes the complex conjugate: for a complex number, \(g = a+bi\), the complex conjugate is: \(\Bar{g} = a-bi\). If \(g\) is a real number, then \(g = \Bar{g}\). For the proof of Eq.~\ref{Kinematic Equation 4}, we get to a form where one of the inner products has an addition in the second term that we have to break up, and no matter which way we rewrite the proof line, one of the inner products ends up with addition in the second term. To proceed, we instead defined the final kinematic equation to hold only for real time. In Lean, this looks like \href{https://atomslab.github.io/LeanChemicalTheories/physics/kinematic_equations.html#real_const_accel'''}{{\Large\texttwemoji{atom symbol}}}:

\begin{alltt}
    \textcolor{indigo}{theorem} \textcolor{maroon}{real_const_accel'''}
    \textcolor{indigo}{(}N : motion_cont_diff_everywhere \(\mathbb{R}\) E\textcolor{indigo}{)}
    \textcolor{indigo}{(}accel_const : N.to_motion.acceleration = λ \textcolor{dgreen}{(}t :\(\mathbb{R}\)\textcolor{dgreen}{)}, A\textcolor{indigo}{)}
    \{n : with_top \(\mathbb{N}\)\}
    :
    
    \(\forall\) t : \(\mathbb{R}\), inner \textcolor{indigo}{(}motion.velocity t\textcolor{indigo}{)} \textcolor{indigo}{(}motion.velocity t\textcolor{indigo}{)} = 
    inner \textcolor{indigo}{(}motion.velocity 0\textcolor{indigo}{)} \textcolor{indigo}{(}motion.velocity 0\textcolor{indigo}{)} + 
    2 * inner A \textcolor{indigo}{(}\textcolor{dgreen}{(}motion.position t\textcolor{dgreen}{)} - \textcolor{dgreen}{(}motion.position 0\textcolor{dgreen}{)}\textcolor{indigo}{)} :=   
\end{alltt}

While we haven't proved that Eq.~\ref{Kinematic Equation 4} doesn't hold for complex time, we encountered difficulties and contradictions when attempting to prove the complex case. Thus, Eq.~\ref{Kinematic Equation 4} currently only holds for real time.

An imaginary-time framework can be used to derive equations of motion from non-standard Lagrangians \cite{popov2005imaginary, rami2013non} to examine hidden properties in classical and quantum dynamical systems in the future. By exploring these proofs in both real and complex time, we illustrate how a proof in one case can be adapted for related cases. Here, four proofs for real numbers can be easily extended to complex numbers by changing the type declared up front, and the validity of the proofs in the more general context is immediately apparent.

\section{Conclusions and Outlook}

\par In this paper, we demonstrate how interactive theorem proving can be used to formally verify the mathematics in science and engineering. We found that, although formalization is slower and more challenging than writing hand-written derivations, our resulting proofs are more rigorous and complete. We observed that in some cases, translating scientific statements into formal language revealed hidden assumptions behind the mathematical derivations. For example, we make explicit common implicit assumptions, such as the denominator must not be zero when we deal with division. As well, in a more abstract way, we have attempted to reveal the formal definitions of equations, such as exactly how pressure is defined as a function or the assumptions of differentiability needed for kinematics. All of these are a result of formalizing these theorems. We concur with others who have discussed the limitations of hand-written proofs and their reliability \cite {bundy2005proof, hales2008formal,avigad2014formally}; formalized proofs can provide greater assurance and robustness.

Importantly, we emphasize that while our proofs are verified to be mathematically correct, this verification does not extend to the external world. This distinction between \emph{syntax} (logical relationships among words and arguments in a language) and \emph{semantics} (whether words are meaningful or arguments are true, according to external reality) in scientific reasoning has been emphasized by logicians such as Alfred Tarski \cite{tarski1944semantic} and Rudolf Carnap \cite{carnap2014logical,carnap1942introduction}. For scientists and engineers, whether a theory is true or meaningful is first and foremost about whether observational data support it -- logical correctness of the derivation is required, of course, but this is typically assumed. Indeed, when one of us described our BET proof to an experimentalist in adsorption, their reply was, "but BET isn't accurate." They knew that BET theory does not \emph{semantically} match experiments in many contexts (in fact, much literature has discussed when BET analysis should \emph{not} be applied, for instance  \cite{ambroz2018evaluation}). BET theory has been a useful conceptual model for the field, but nonetheless relies on approximations that often drift far from reality. In this work, we only claim to rigorously establish the \emph{syntax} of the theories we describe. Nonetheless, Lean operating with input/output functions can receive data from the external world, which may open possibilities for \emph{semantically} grounding its logical conclusions in certain contexts, as well.

The Lean Theorem Prover is especially powerful, as it facilitates the re-use of theorems and the construction of higher-level mathematical objects from lower-level ones. We showed how this feature can be leveraged in science proofs; after a fundamental theory is formally verified, it can then be used in the development of other theories. This can be approached in two ways: \emph{definitions} can be directly reused in subsequent proofs, and \emph{structures} can enable hierarchies of related concepts, from general to more specific. Thus we have not just proved a few theorems about scientific objects but have begun to create an interconnected structure of formally verified proofs relating fields of science. 

While learning Lean and writing the proofs appearing here, we routinely asked ourselves, ``How \emph{do} I close this goal? I wish there was a way to automate this.'' In fact, the first vision for computer-assisted proofs in the 1950s and 60s was to automate the process fully \cite{harrison2014history}; \emph{interactive} theorem provers that ``merely'' check human-written proofs didn't appear until later.
But historically, automated theorem provers (ATPs) made progress on narrow classes of problems (e.g., problems in first-order logic \cite{kovacs2013first}) but couldn't address proofs in advanced math (except when problems are described in such simple terms, like the Robbins Conjecture \cite{Mccune_1997}).
In short, theorem proving is like searching for a path from premises to conjecture, but in a realm with an ``infinite action space'' \cite{polu2022formal} -- traditional algorithms have been inadequate.
For complex proofs, interactive theorem provers (ITPs) have been more successful, because they facilitate human creativity in writing proofs, while leveraging the rigor of the computer for checking them and providing feedback to the user.
Modern ITPs also use the computer for small-scale automation via tactics; the human provides strategy while the computer executes tactics.
Complex tactics sometimes blur the line between automated and interactive theorem proving. 
For example, Isabelle (an ITP) has the Sledgehammer tactic \cite{blanchette2011automatic}, which takes the current proof state and attempts to transform it into an equivalent problem in first-order logic, which can then be efficiently solved using an ATP. 

Recent approaches have leveraged machine learning to expand the capabilities of automated theorem proving. Theorem proving can be framed as a reinforcement learning problem \cite{kaliszyk2018reinforcement, crouse2021deep}, in which an agent is to learn an effective theorem proving policy via rewards from successfully proving theorems. 
``Autoformalization'' refers to the translation of informal proofs into formal proofs, akin to translating text from one language to another (but with extremely strict requirements on the formal side) \cite{szegedy2020}.
Theorem proving can also be framed as a next-word-prediction problem (``auto-complete'' for math proofs) in which a database of formal math proofs is used to train a language model to predict the next word in the proof.
Large language models (LLMs) like ChatGPT \cite{brown2020language, chen2021evaluating} have some emergent reasoning abilities \cite{ wei2022chain} but often make mistakes and cannot be trusted.
By connecting language models with ITPs to provide feedback, training them on proof databases like \texttt{mathlib}, and deploying them as part of traditional search algorithms, progress has been made toward automating proofs in Lean \cite{proofartifact2021, GPT-f_lean, yang2023leandojo}, even to the point of generating correct solutions to International Math Olympiad problems \cite{poluformal2022}.

This interplay between creative but unreliable generative algorithms and the strict logic of a proof-checking system may be a model for future AI-driven discovery in science, especially for discovering new theories. 
An early example of this is AI-Descartes, in which a symbolic regression algorithm generates equations to match experimental data, which is then combined with an automated theorem prover to establish the equations' ``derivability'' with respect to a scientific theory \cite{cornelioCombiningDataTheory2023}.
However, in this work, each theory required human expertise to be expressed in formal language, and reliance on an automated theorem prover limited the scope of theories to those expressible in first-order logic.
AI tools that can autoformalize the informal scientific literature, generate novel theories, and auto-complete complex proofs could open new avenues for automating theory discovery.
LLMs have demonstrated capabilities in solving chemistry problems \cite{ hocky2022natural, white2022large}, as well as answering scientific question-and-answer problems invoking quantitative reasoning \cite{10.48550/ARXIV.2206.14858}. However, LLMs are unreliable - they famously ``hallucinate'' (generate falsehoods) and are biased or unreliable evaluators of their own outputs \cite{bran2023chemcrow,liu2023gpteval}.
Pairing them with external tools \cite{autoGPT, langchain, openAIfunc} improves their capabilities; theorem provers could play a role like that.
How will these models be trained? We suggest two avenues: training on human-written databases of formal proofs in science and engineering (which are yet to be written) and leveraging interactive feedback from Lean through tools like LeanDojo \cite{yang2023leandojo}. Beyond being formally grounded in axiomatic mathematics, formal proofs in science and engineering are machine-readable instances of correct mathematical logic that could serve as a foundation for artificial intelligences aiming to learn, reason, and discover in science \cite{ bradshaw1983studying, kitano2021nobel, krenn2022scientific}.

Our next goals are to continue building out classical thermodynamics, formalize statistical mechanics, and eventually construct proofs relating the two fields. We are also interested in laying the foundations for classical mechanics in Lean and formalizing more difficult proofs like Noether's Theorem \cite{kosmann2011noether} (a basis for deriving conservation laws) or establishing the 2nd Law of Thermodynamics axiomatically \cite{lieb1999physics}.

The proofs in this paper were written in Lean 3 \cite{avigad2014formally}, because the extensive \texttt{mathlib} library was only available in Lean 3 when we began. While Lean 3 was designed for theorem proving and management of large-scale proof libraries, the new version, Lean 4 \cite{moura2021lean}, is a functional programming language for writing proofs and programs, as well as proofs about programs \cite{moura2021lean, ProgLean}. The Lean community recently finished porting \texttt{mathlib} to Lean 4; we recommend future proofs be written in Lean 4, which is more capable, versatile, and easy to use compared to Lean 3. With Lean 4, we are bridging formally-correct proofs with executable functions for bug-free scientific computing; we will be elaborating on that in future work.

We hope these expository proofs in adsorption, thermodynamics, and kinematics will inspire others to consider what proofs and derivations could be formalized in their fields of expertise. Virtually all mathematical concepts can be established using dependent type theory; the density functionals, partial derivatives, N-dimensional integrals, and random variables appearing in our favorite theories \emph{should be expressible in Lean}.
Just as an ever-growing online community of mathematicians and computer scientists is building \texttt{mathlib} \cite{zulipLean}, we anticipate a similar group of scientists building a library of formally-verified scientific theories and engineering mathematics. To join, start learning Lean, join the online community, and see what we can prove!

\section*{Author Contributions}

We summarize author contributions using the CRediT system. Conceptualization: TRJ; 
Data curation: MPB, PF, SS; Formal Analysis: MPB, SS, PF, AHD, CMW, TRJ; Funding Acquisition: TRJ; Methodology: MPB, TRJ; Project administration: MPB, SS, TRJ; Software: MPB, SS, TRJ; Supervision: TRJ; Validation: Lean; Visualization: MPB, SS, PF, TRJ; Writing, Original draft: MPB, SS, PF, TRJ; Writing, review and editing: MPB, SS, PF, TRJ.

\section*{Acknowledgements}
We are grateful to the Lean prover community and contributors of \texttt{mathlib} on whose work this project is built. We especially thank Kevin Buzzard, Patrick Massot, Tomas Skrivan, Eric Wieser, and Andrew Yang for helpful comments and discussions around our proof structure and suggestions for improvement. We thank Charles Fox, Mauricio Collares, and Ruben Van de Velde for helping with the website. We thank two anonymous peer reviewers, as well as Rose Bohrer, John Keith, and Ben Payne for reading the manuscript and providing helpful feedback. This material is based upon work supported by the National Science Foundation under Grant No. (NSF \#2138938), as well as startup funds from the University of Maryland, Baltimore County.

\raggedbottom
\section*{Supporting Information}
The supporting information provides additional background on how Lean works (Section \ref{LeanAdditionalInfo}) and all the additional proofs (Section \ref{SIproofs}), including an improved version of Langmuir adsorption model (\ref{LangmuirSI}), the final derived form of BET adsorption model (\ref{BETSI}), and the antiderivative proofs (\ref{antiderivSI}) that was used for kinematic equations. All code and proofs for this project are available in our GitHub repository \href{https://atomslab.github.io/LeanChemicalTheories/}{{\Large\texttwemoji{atom symbol}}}.

\section*{Conflicts of Interest}
The authors declare that they have no known competing financial interests or personal relationships that could have appeared to influence the work reported in this paper.
\raggedbottom
\newpage

\section*{Glossary of Mathematical Terms and Symbols}\label{Glossary}
\begin{table}[H]
\begin{tabular}{p{0.1\textwidth} p{0.42\textwidth} p{0.41 \textwidth}}
\hline
\textbf{Term}        & \textbf{Definition}                                                                                                                            & \textbf{Example}                                                                                                                                    \\ \hline
Axiom                & A self-evident truth which is assumed to be true and doesn’t require proof.                                                                    & Two sets are equal if they have the same elements; this is not proved, it is assumed.                                                            \\
Theorem              & A theorem is a proposition or statement in math that can be demonstrated to be true by accepted mathematical operations and arguments.         & The Pythagorean theorem, $a^2 + b^2 = c^2$ for all right triangles                                                                                            \\
Lemma                & A true statement which is used as a stepping stone to prove other true statements. A lemma is a smaller, less important result than a theorem. & All numbers multiplied by 2 are even. Proving this could be an intermediate result used for other proofs.                                                               \\
Proposition          & A true or false statement.                                                                                                                     & Socrates is mortal, all swans are white, and $3 < 4$ are propositions.   \\
Hypothesis (Premise) & A statement assumed to be true that the proof follows from. It can also be thought of as the conditions or prerequisites for the theorem to hold. We emphasize that the math community uses "hypothesis" somewhat differently than the scientific community.                      & If all four sides of a rectangle have the same length, it is a square. The hypotheses would be ``the shape is a rectangle (has four sides and four equal, right angles)'' and ``all sides have the same length.'' \\
Conjecture           & A statement which is proposed to be true, but no proof has been found yet.                                                                     & Goldbach's conjecture: every even number greater than 2 is the sum of two prime numbers. This hasn't been proven true or false yet.                 \\
Proof                & A sequence of logical steps which conclude that a statement is true from its hypotheses.                                                       & The proof of the Pythagorean theorem using geometry                                                                                                 \\
Function             & An expression that defines a relation between a set of inputs and a set of outputs.                                                            & \(f(x)=x^2\) relates (or maps) the set of real numbers $x$ to their square.                                                            \\
Tactic             & A command used to construct or manipulate proofs. Tactics in Lean provide a way to automate certain proof steps or apply predefined proof strategies to make the process of constructing formal proofs more efficient and convenient.  & \texttt{rw} is "rewrite", a simple tactic that performs substitution for equalities. \texttt{ring} is a more complex tactic for automatically closing goals requiring numerous algebraic operations, without the user specifying all the steps.  \\
Type                 & A type can be thought of as a set, or category, that contains terms. In other programming languages, types define the category of data certain objects have (e.g. floats, strings, integers). Types in Lean work this way, too, and have more features: they can depend on values, as well as be the subject of proofs.          & The natural numbers are a type. The Booleans (True and False) are also a type. Functions from integers to reals are also a type.                                                            \\
Term                 & Terms are members of a type.                                                                                                                   & Considering the type of natural numbers, then numbers like 1, 2, 3, and 8 are terms of that type.    
   \\
\(\mathbb{N}\)                 & Symbol for the set of natural numbers                                                                                                                   & The numbers 0, 1, 2, 3, 4...   
   \\
\(\mathbb{Z}\)                   & Symbol for the set of integers                                                                                                                   & The numbers, -3, -2, -1, 0, 1, 2 ...
   \\
\(\mathbb{Q}\)                   & Symbol for the set of rational numbers                                                                                                                  & The numbers, \(\frac{1}{2}, \frac{3}{4}, \frac{5}{9}\), etc. 
\\
\(\mathbb{R}\)                 & Symbol for the set of real numbers                                                                                                                   & -1, 3.6, Euler's number, $\pi$, $\sqrt{2}$, etc.
   \\
\(\mathbb{C}\)                 & Symbol for the set of complex numbers                                                                                                                   & -1, 5 + 2$i$, $\sqrt(2) + 5i$, etc.
   \\
\(\forall\)                 & Logical symbol for "for all"                                                                                                                   & 
   \\
\(\exists\)                 & Logical symbol for "there exists"                                               &                                                                    
\end{tabular}

\end{table}

\bibliographystyle{vancouver}
\bibliography{references} 
\newpage
\section{Supporting Information}
 
\subsection{Additional Background}\label{LeanAdditionalInfo}
Lean is an open source theorem prover developed by Microsoft Research and Carnegie Mellon University, based on dependent type theory, with the goal to formalize theorems in an expressive way \cite{de_moura_kong_avigad_van_doorn_von_raumer_2015}. Lean supports user interaction and constructs axiomatic proofs through user input, allowing it to bridge the gap between interactive and automated theorem proving. Like Mizar \cite{rudnicki1992overview} and Isabelle \cite{wenzel2002isabelle}, Lean allows user to state definitions and theorems but also combines more imperative tactic styles as in Coq \cite{barras1997coq}, HOL-Light \cite{gordon1993introduction}, Isabelle \cite{nipkow2002isabelle}  and PVS \cite{owre1992pvs} to construct proofs. The ability to define mathematical objects, rather then just postulate them is where Lean gets its power \cite{avigad2015theorem}. It can be used to create an interconnected system of mathematics where the relationship of objects from different fields can be easily shown without loosing generality.

\setcounter{figure}{0}
\renewcommand{\figurename}{Figure}
\renewcommand{\thefigure}{S\arabic{figure}}

\begin{figure}[h]
\centering
\includegraphics[width=1.0\textwidth]{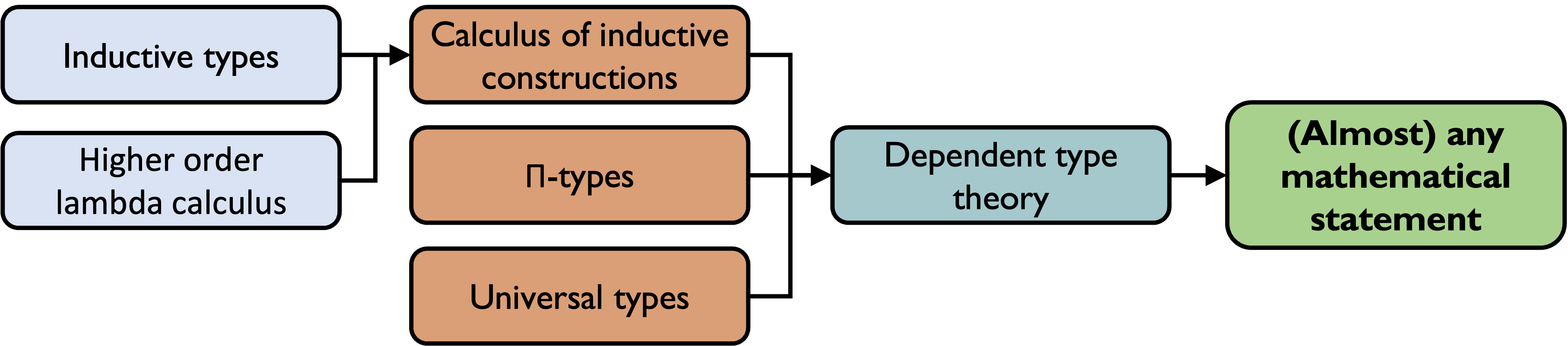}
\caption{Overview of Lean Theorem Prover}
\label{leanoverview}
\end{figure}

\par As mentioned above, the power of Lean comes from the ability to define objects and prove properties about them. In Lean, there are three ways to define new Types: type universes, Pi types, and inductive types. The first two are used to construct the basis of dependent type theory, and are used for more theoretical, foundational stuff. Instead we will focus on the use of inductive types. Standard inductive types, known as just \textit{inductive types}, are built from a set of constructors and well found \textit{recursion}. Non-recursive inductive types that contain only one constructor are called \textit{structures}. 
\par Many mathematical objects in Lean can be constructed through inductive types, which is a type built from a set of constructors and proper recursion \cite{dybjer1994inductive}. The natural numbers are an inductive type, defined using Peano's Encoding \cite{skolem1955peano}. This requires two constructors, a constant element, 0 : nat, and a function called the successor function, S. Then one can be constructed as S(0), two can be constructed as S(S(0)), etc. 
%\newpage

In Lean, the natural numbers are defined as:

\begin{alltt}
    \textcolor{indigo}{inductive} \textcolor{maroon}{nat}
    | zero : nat
    | succ \textcolor{indigo}{(}n : nat\textcolor{indigo}{)} : nat
\end{alltt}

Here, the type \textit{nat} is defined through recursion by a constant element, zero, and a function. With this, the \textit{def} command is used to define properties about the class, like addition or multiplication. For instance, the addition of the natural numbers is defined as:

\begin{alltt}
    \textcolor{dgreen}{protected} \textcolor{indigo}{def} \textcolor{maroon}{add} : nat → nat → nat
    | a  zero     := a
    | a  \textcolor{indigo}{(}succ b\textcolor{indigo}{)} := succ \textcolor{indigo}{(}add a b\textcolor{indigo}{)}
\end{alltt}

Addition is defined as a function that takes in two natural numbers and outputs a natural number. Since the natural numbers are created from two constructors, there are two cases of addition that must be shown. The first is a general natural number plus zero which yields the general natural number, and the next is a general natural number plus the successor of a general natural number. The second case used recursion and calls \textit{add} again until it reduces to zero.

\par The other way to define types is using \textit{structure} which allows us to add constraints to a type variable. For instance, the \textit{class} \textit{has\_add} constrains a type to have a function called \textit{add} which represents addition.

\begin{alltt}
    \textcolor{indigo}{class} \textcolor{maroon}{has_add} \textcolor{indigo}{(}\(\alpha\) : Type u\textcolor{indigo}{)} := 
    \textcolor{indigo}{(}add : \(\alpha\) → \(\alpha\) →  \(\alpha\)\textcolor{indigo}{)}
\end{alltt}

This can be used for more advanced ideas, like defining rings or abelian groups. We can use class to define areas of science as new types with constraints to follow certain rules.

\subsection{Additional Proofs} \label{SIproofs}
\subsubsection{Langmuir Adsorption}\label{LangmuirSI}

The first Langmuir proof introduced earlier states every premise explicitly but however we can condense that by rewriting \textit{hrad} and \textit{hrd} into \textit{hreaction} to yield \textit{k\_ad*Pₐ*S = k\_d*A} and we can then rewrite \textit{h\(\theta\)} and \textit{hK} in the goal statement. While \textit{hrad}, \textit{hrd}, \textit{h\(\theta\)}, and \textit{hK} have scientific significance, they do not have any mathematical significance. In Lean it looks like:

\begin{alltt}
    \textcolor{indigo}{theorem} \textcolor{maroon}{Langmuir_single_site2}
    \textcolor{indigo}{(}Pₐ k_ad k_d A S: \textcolor{tpurple}{\(\mathbb{R}\)}\textcolor{indigo}{)}
    \textcolor{indigo}{(}hreaction : k_ad*Pₐ*S = k_d*A\textcolor{indigo}{)}
    \textcolor{indigo}{(}hS : S \(\ne\) 0\textcolor{indigo}{)}
    \textcolor{indigo}{(}hk_d : k_d \(\ne\) 0\textcolor{indigo}{)}
    : A/\textcolor{indigo}{(}S+A\textcolor{indigo}{)} = k_ad/k_d*Pₐ/\textcolor{indigo}{(}1+k_ad/k_d*Pₐ\textcolor{indigo}{)} :=
\end{alltt}
\par However, while those four variables do not have any mathematical significance, and only serve to hinder our proofs, they do have scientific significance, and we do not want to just omit them. Instead we can use the \textit{let} command to create an in-line, local definition. This allows us to have the applicability of the theorem, while still having scientifically important variables. In Lean, this looks like \href{https://atomslab.github.io/LeanChemicalTheories/adsorption/langmuir_kinetics.html} {{\Large\texttwemoji{atom symbol}}}:

\begin{alltt}
    \textcolor{indigo}{theorem} \textcolor{maroon}{Langmuir_single_site}
    \textcolor{indigo}{(}Pₐ k_ad k_d A S : \textcolor{tpurple}{\(\mathbb{R}\)}\textcolor{indigo}{)}
    \textcolor{indigo}{(}hreaction : \textcolor{indigo}{let} r_ad := k_ad*Pₐ*S, r_d := k_d*A \textcolor{indigo}{in} r_ad = r_d\textcolor{indigo}{)} 
    \textcolor{indigo}{(}hS : S \(\ne\) 0\textcolor{indigo}{)}
    \textcolor{indigo}{(}hk_d : k_d \(\ne\) 0\textcolor{indigo}{)}
    : 
    \textcolor{indigo}{let} \(\theta\) := A/(S+A),
        K := k_ad/k_d \textcolor{indigo}{in} 
    \(\theta\) = K*Pₐ/\textcolor{indigo}{(}1+K*Pₐ\textcolor{indigo}{)} :=
\end{alltt}

The first line after the \textit{theorem} statement, gives the variables use in the proof. Notice that \(r_{ad}\), \(r_d\), \(K_{eq}\), and \(\theta\) are not defined as variables. Instead, the \textit{let} statement defines those four variables in their respective premise or goal. Then, in the proof we can simplify the \textit{let} statement to get local definitions of those variables, just like \textit{hrad}, \textit{hrd}, \textit{h\(\theta\)}, and \textit{hK}. While this version of proof follow the same proof logic minus the two initial rewrites from earlier version, however if we stick with the first proof, we will find it very difficult to use compared to using this proof above, because of all those hypothesises. Suppose we wanted to prove \textit{langmuir\_single\_site2} and we already have proven \textit{langmuir\_single\_site}. We would find it impossible to use \textit{langmuir\_single\_site} because we are missing premises like \textit{hrad} or \textit{hrd}. Yet, we could prove the other way, ie. use \textit{langmuir\_single\_site} to prove \textit{langmuir\_single\_site2}. Having all of those extra premises that define the relation between variables only serves to hinder the applicability of our proofs.

\subsubsection{BET Adsorption}\label{BETSI}

\par We continue the derivation of Equation 27 from the paper that aims to redefine \textit{x} as \(x = P/P_0\), by recognizing that the volume should approach infinity at the saturation pressure, and, mathematically, it approaches infinity as x approaches one from the left. For x to approach one, pressure must approach \(1/C_L\). First, we show that Equation 26 from the paper approaches infinity as \textit{P} approaches \(1/C_L\). We specifically require it to approach from the left because volume approaches negative infinity if we come from the right. In Lean, this looks like \href{https://atomslab.github.io/LeanChemicalTheories/adsorption/BETInfinite.html#tendsto_at_top_at_inv_CL}{{\Large\texttwemoji{atom symbol}}}:

\begin{alltt}
    \textcolor{indigo}{lemma} \textcolor{maroon}{BET.tendsto_at_top_at_inv_CL}
    : filter.tendsto \textcolor{maroon}{brunauer_26}
    \textcolor{indigo}{(}nhds_within \textcolor{dgreen}{(}1/C_L\textcolor{dgreen}{)} \textcolor{dgreen}{(}set.Ioo 0 \textcolor{iibrown}{(}1/C_L\textcolor{iibrown}{)}\textcolor{dgreen}{)}\textcolor{indigo}{)} 
    filter.at_top:=
\end{alltt}

The function \textit{filter.tendsto} is the generic definition of the limit. It has three inputs, the function, what the independent variable approaches, and what the function approaches, in that order. We split this into three lines to better visualize what is happening. First, we are using the object \textit{brunauer\_26}, which is the BET equation as a function of pressure in terms of x. Next, \textit{(nhds\_within (1/C\_L) (set.Ioo 0 (1/C\_L)))} is how we say approaches \(1/C_L\) from the left. \textit{nhds\_within} means the intersection of a neighborhood, abbreviated as \textit{nhds}, and a set. A neighborhood of a point is the open set around that point. \textit{set.Ioo} designates a left-open right-open interval. Here we have the interval \((0,1/C_L)\). The intersections of the neighborhood and this set constrains us to approach the neighborhood from the left. The final part is \textit{filter.at\_top} which is a generalization of infinity, and just says our function approaches infinity. 
\par In the original derivation done by Brunauer et al, they wish to show that \(P_0 = 1/C_L\) because as pressure approaches each of these values, volume approaches infinity, these two values are equal. It should be noted that this idea is only true if \textit{C}, the BET constant, is greater than or equal to one. If not, the function has two points where it hits infinity in the positive pressure region. We also have problems showing the congruence of such a fact in Lean, since such a relation has yet to be formalized and the congruence of two \textit{nhds\_within} has not been shown. For now, we use the lemma above to prove a simplier version of the theorem where we assume \(P_0=1/C_L\), and show that with this assumption, \(V\) approaches infinity. In Lean, this looks like \href{https://atomslab.github.io/LeanChemicalTheories/adsorption/BETInfinite.html#brunauer_27}{{\Large\texttwemoji{atom symbol}}}:

\begin{alltt}
    \textcolor{indigo}{theorem} \textcolor{maroon}{brunauer_27}
    \textcolor{indigo}{(}h1 : P_0 = 1/C_L\textcolor{indigo}{)}
    : filter.tendsto \textcolor{maroon}{brunauer_26}  \textcolor{indigo}{(}nhds_within \textcolor{iibrown}{(}P_0\textcolor{iibrown}{)} \textcolor{dgreen}{(}set.Ioo 0 \textcolor{iibrown}{(}P_0\textcolor{iibrown}{)}\textcolor{dgreen}{)}\textcolor{indigo}{)} filter.at_top:=
\end{alltt}

The proof of this theorem involves rewriting h1, and then applying the lemma proved above. While we would prefer to prove that \(P_0 = 1/C_L\), this proof will serve as a placeholder, until Mathlib builds out more math related to the congruence of this subject. This theorem does not use a lcoal definition, like Langmuir, because \(P_0\) is already defined as a variable using \textit{constant}.
\par Finally, we formalize the derivation of Equation 28 from the paper, givne by Equation \ref{BET 28}.

\begin{equation}
    \label{BET 28}
     \frac{V}{A*V_0} = \frac{CP}{(P_O-P)(1+(C-1)(P/P_0)}
\end{equation}
Just like Equation \ref{BET 26}, we first define Equation \ref{BET 28} at an object then formalize the derivation of this object. In Lean, the object looks like \href{https://atomslab.github.io/LeanChemicalTheories/adsorption/BETInfinite.html#brunauer_28}{{\Large\texttwemoji{atom symbol}}}:

\begin{alltt}
    \textcolor{indigo}{def} \textcolor{maroon}{brunauer_28} := \(\lambda\) P : \(\mathbb{R}\), C*P/\textcolor{indigo}{(}\textcolor{dgreen}{(}P_0-P\textcolor{dgreen}{)}*\textcolor{dgreen}{(}1+\textcolor{iibrown}{(}C-1\textcolor{iibrown}{)}*\textcolor{iibrown}{(}P/P_0\textcolor{iibrown}{)}\textcolor{dgreen}{)}\textcolor{indigo}{)} 
\end{alltt}

Now we can prove a theorem that formalizes the derivation of this object \href{https://atomslab.github.io/LeanChemicalTheories/adsorption/BETInfinite.html#brunauer_28_from_seq}{{\Large\texttwemoji{atom symbol}}}: 

\begin{alltt}
    \textcolor{indigo}{theorem} \textcolor{maroon}{brunauer_28_from_seq}
    \textcolor{indigo}{\{}P V_0: \textcolor{tpurple}{\(\mathbb{R}\)}\textcolor{indigo}{\}}
    \textcolor{indigo}{(}h27 : P_0 = 1/C_L\textcolor{indigo}{)}
    \textcolor{indigo}{(}hx1: \textcolor{dgreen}{(}x P\textcolor{dgreen}{)} < 1\textcolor{indigo}{)}
    \textcolor{indigo}{(}hx2 : 0 <  \textcolor{dgreen}{(}x P\textcolor{dgreen}{)}\textcolor{indigo}{)}
    : let Vads :=  V_0 * \(\sum'\) \textcolor{indigo}{(}k : \(\mathbb{N}\)\textcolor{indigo}{)}, ↑k * \textcolor{indigo}{(}seq P k\textcolor{indigo}{)},
          A :=  \(\sum'\) \textcolor{indigo}{(}k : \(\mathbb{N}\)\textcolor{indigo}{)}, \textcolor{indigo}{(}seq P k\textcolor{indigo}{)} in
      Vads/A = V_0*\textcolor{indigo}{(}\textcolor{maroon}{brunauer_28} P\textcolor{indigo}{)} :=
\end{alltt}
Rather then explicitly solving the sequence ratio, like we did for Equation \ref{BET 26}, we can now use the theorem that derived Equation \ref{BET 26} to solve the left hand side of our new goal. We then have a goal where we show that Equation \ref{BET 28} is just a rearranged version of Equation \ref{BET 26}, which is done through algebraic manipulation. 

\subsubsection{The antiderivative in Lean} \label{antiderivSI}
\par For a function, \(f\), the antiderivative of that function, given by \(F\), is a differentiable function, such that the derivative of \(F\) is the original function \(f\). In Lean, we formalize the general antiderivative and show how it can be used for several specific applications, including the antiderivative of a constant, of a natural power, and of an integer power. We generalize our functions as a function from a general field onto a vector field, \(f : \mathbb{K} \rightarrow E\). This allows us to apply the theorems to any parametric vector function, including scalar functions. 
\par Our goal is to show, from the assumption that \(f(t)\) is the derivative of \(F(t)\) and \(f(t)\) is the derivative of \(G(t)\), then we have an equation \(F(t) = G(t) + F(0)\), which is the antiderivative of \(f(t)\). \(G(t)\) is the variable portion of the equation. For example, if the antiderivative is of the form \(F(t) = t^3 + t + 6\), then \(G(t) = t^3 + t\) and \(F(0) = 6\). \(F(0)\) is the constant of integration, but written in a more explicit relation to the function. Since \(G(t)\) is the function of just variables, we have as another premise \(G(0) = 0\). 
\par The first goal is to show that a linearized version of the antiderivative function holds. We can rewrite \(F(t)\) so that is linear by moving \(G(t)\) to the left hand side, leaving us with an equation that equals a constant.
\begin{equation}
    \label{Constant Function}
    F(t)-G(t) = C
\end{equation}
Thus, we can relate any two points along this function, \(\forall x y, F(x)-G(x)=F(y)-G(y)\). To show this holds, we recognize that if Equation \ref{Constant Function} is constant, then the derivative of this function is equal to zero.
\begin{equation}
    \label{Deriv of Constant}
    \frac{d}{dt}(F(t)-G(t)) = 0
\end{equation}
Next, we apply the linearity of differentiation to Equation \ref{Deriv of Constant} to get a new form: \(\frac{d}{dt}F(t) - \frac{d}{dt}G(t) = 0\), and rearrange to get:
\begin{equation}
    \label{Antideriv sub}
    \frac{d}{dt}F(t) = \frac{d}{dt}G(t)
\end{equation}
From the first premise, we assumed that \(f(t)\) is the derivative of \(F(t)\). From our second premise, we assumed that \(f(t)\) is also the derivative of \(G(t)\). Thus, applying both premises, we can simplify Equation \ref{Antideriv sub} to:
\begin{equation}
    f(t) = f(t)
\end{equation}
which we recognize to be correct. 
\par Now that we have a new premise to use, given by Equation \ref{Antideriv Const}, we can specialize this function to get our final form.
\begin{equation}
    \label{Antideriv Const}
    \forall x y, F(x)-G(x)=F(y)-G(y)
\end{equation}
We specialize the universals by supplying two old names. For x, we use t (the variable we have been basing our differentiation around), and for y we use 0. Thus, Equation \ref{Antideriv Const} becomes:
\begin{equation}
    \label{Antideriv Const2}
     F(t) - G(t) = F(0) - G(0)
\end{equation}
Our third premise was that \(G(0) = 0\), so we can simplify and rearrange Equation \ref{Antideriv Const}, to get our final form:
\begin{equation}
    \label{Antideriv Final Form}
    F(t) = G(t) + F(0)
\end{equation}
Which satisfies the goal we laid out in the beginning. In Lean, the statement of this theorem looks like \href{https://atomslab.github.io/LeanChemicalTheories/math/antideriv.html#antideriv}{{\Large\texttwemoji{atom symbol}}}:
\begin{alltt}
    \textcolor{indigo}{theorem}  \textcolor{maroon}{antideriv} 
    \textcolor{indigo}{\{}\textcolor{tpurple}{E} : Type u_2\textcolor{indigo}{\}}  \textcolor{indigo}{\{}\textcolor{tpurple}{\(\mathbb{K}\)}: Type u_3\textcolor{indigo}{\}} \textcolor{indigo}{[}is_R_or_C \textcolor{tpurple}{\(\mathbb{K}\)}\textcolor{indigo}{]} \textcolor{indigo}{[}normed_add_comm_group \textcolor{tpurple}{E}\textcolor{indigo}{]} 
    \textcolor{indigo}{[}normed_space \textcolor{tpurple}{\(\mathbb{K}\)} \textcolor{tpurple}{E}\textcolor{indigo}{]}
    \textcolor{indigo}{\{}f F G: \textcolor{tpurple}{\(\mathbb{K}\)} → \textcolor{tpurple}{E}\textcolor{indigo}{\}} \textcolor{indigo}{(}hf : \(\forall\) t, has_deriv_at F \textcolor{dgreen}{(}f t\textcolor{dgreen}{)} t\textcolor{indigo}{)} 
    \textcolor{indigo}{(}hg : \(\forall\) t, has_deriv_at G \textcolor{dgreen}{(}f t\textcolor{dgreen}{)} t\textcolor{indigo}{)}
    \textcolor{indigo}{(}hg' : G 0 = 0\textcolor{indigo}{)}
    : F = \(\lambda\) t, G t + F(0) :=
\end{alltt}
Applying the \textit{antideriv} theorem to examples is very straight forward. We will show an example by deriving the antiderivative of a constant function. In Lean, we would state this as \href{https://atomslab.github.io/LeanChemicalTheories/math/antideriv.html#antideriv_const}{{\Large\texttwemoji{atom symbol}}}:
\begin{alltt}
    \textcolor{indigo}{theorem} \textcolor{maroon}{antideriv_const}
    \textcolor{indigo}{(}F : \textcolor{tpurple}{\(\mathbb{K}\)} → E\textcolor{indigo}{)} {k : \textcolor{tpurple}{E}}
    \textcolor{indigo}{(}hf : \(\forall\) t, has_deriv_at F k t\textcolor{indigo}{)}:
    \textcolor{indigo}{(}F = λ \textcolor{dgreen}{(}x : \textcolor{tpurple}{\(\mathbb{K}\)})\textcolor{dgreen}{)}, x\(\cdot\)k + F 0\textcolor{indigo}{)} :=
\end{alltt}
Here we say that the derivative of \(F(x)\) is the constant \(k\), and want to show that \(F(x) = x\cdot k + F(0)\), where the "\(\cdot\)" operator stands for scalar multiplication. To use the \textit{antideriv} theorem, we must show that its premises follow, meaning we must show:
\begin{alltt}
    \(\forall\) t, has_deriv_at F k t
    \(\forall\) t, has_deriv_at x\(\cdot\)k k t
    0\(\cdot\)k = 0
\end{alltt}
The first goal is explicitly given in our premises, \textit{hf}. The next goal can be derived by taking out the constant, and showing that the function \(x\) has a derivative equal to \(1\). The final goal can be easily proven by recognizing zero multiplied by anything is zero. Thus, we have formalized antiderivative of a constant function, and can use this same process for any other function. The antiderivative is especially important for deriving the kinematic equations, as seen in the next section.

\end{document}